\documentclass[12pt,preprint]{aastex}

\usepackage{color}

\begin{document}
\title{Compact Remnant Mass Function:  Dependence on the Explosion 
Mechanism and Metallicity}

\def\msun{\,{\rm M}_{\odot}}
\def\rsun{{\rm ~R}_{\odot}}
\def\myr{{\rm ~Myr}}
\def\mdot{\dot M}
\def\mpy{{\rm ~M}_{\odot} {\rm ~yr}^{-1}}

\author{Chris L. Fryer\altaffilmark{1,2,3},Krzysztof
Belczynski\altaffilmark{4,5}, Grzegorz Wiktorowicz\altaffilmark{4}, 
Michal Dominik\altaffilmark{4},Vicky Kalogera\altaffilmark{6}, 
Daniel E. Holz\altaffilmark{7,8}}

\altaffiltext{1}{CCS Division, Los Alamos National Laboratory, 
Los Alamos, NM 87545}
\altaffiltext{2}{Department of Physics, The University of Arizona,
Tucson, AZ 85721} 
\altaffiltext{3}{Department of Physics and Astronomy, The University of New Mexico,
Albuquerque, NM 87131} 
\altaffiltext{4}{Astronomical Observatory, University of Warsaw, Al. Ujazdowskie 4, 00-478 Warsaw, Poland}
\altaffiltext{5}{Center for Gravitational Wave Astronomy, University of Texas at Brownsville, TX 78520, USA}
\altaffiltext{6}{Center for Interdisciplinary Exploration and Research in Astrophysics (CIERA) \& Dept.\ of Physics and Astronomy, Northwestern University, 2145 Sheridan Rd., Evanston, IL 60208}
\altaffiltext{7}{Theory Division, Los Alamos National Laboratory, 
Los Alamos, NM 87545}
\altaffiltext{8}{Enrico Fermi Institute, Department of Physics, and Kavli Institute for Cosmological Physics, University of 
Chciago, Chicago, IL 60637}

\begin{abstract}

The mass distribution of neutron stars and stellar-mass black holes
provides vital clues into the nature of stellar core collapse and the
physical engine responsible for supernova explosions.  Using
recent advances in our understanding of supernova engines,
we derive mass distributions of stellar compact remnants.  We provide 
analytical prescriptions for compact object masses for major 
population synthesis codes.  In an accompanying paper, Belczynski et al., 
we demonstrate that these qualitatively new results for compact objects can 
explain the observed gap in the remnant mass distribution between 
$\sim 2-5$\,M$_\odot$ and that they place strong constraints on the 
nature of the supernova engine.  Here, we show that advanced 
gravitational radiation detectors (like LIGO/VIRGO or the Einstein 
Telescope) will be able to further test the supernova explosion 
engine models once double black hole inspirals are detected.

\end{abstract}


\section{Introduction}

Neutron stars (NSs) and black holes (BHs) are among the most exotic
objects produced in nature.  They are formed in the core collapse of
massive stars and, in many cases, their formation is associated with
powerful astrophysical transients such as supernovae and gamma-ray
bursts.  By studying the masses of these objects we can better
understand their formation process and associated explosions.  In
addition, accurate measurements of masses of NSs and BHs provide
essential input to our understanding of a wide range of astrophysical
phenomena produced by these objects, from gravitational waves formed in
compact object mergers to X-ray bursts.

Observations of binaries containing NSs and BHs
(e.g.  X-ray binaries, X-ray bursts, and binary pulsars) place
constraints on their mass distribution.  Estimates of the NS
mass distribution have benefited from observations of close pulsar
binary systems where extremely accurate masses can be obtained through
pulsar timing.  Originally, analyses of these binaries suggested a
very narrow mass distribution around 1.35\,M$_\odot$ (e.g. Thorsett \&
Chakrabarty 1999). More recently, as
available data have increased and become more refined, it has become
clear that the mass distribution is at least bimodal, and likely has a
wide spread ranging from low masses up to the maximum NS
mass limit \citep{kaper06,Freire08,Nice08,van07,Sch10,Kiz11}.  
The observation of a $\sim2.0$\,M$_\odot$ NS~\citep{Dem10} 
is an indication of the width of the distribution.

Black-hole mass measurements rely on a complex combination of
challenging observations of X-ray binaries (in quiescence, if they are
transient) and of modeling of photometric and spectroscopic data. The
uncertainties associated with these measurements are more significant
than in the case of NSs. Early analysis
~\citep{bailyn98} argued that the measurements are consistent with a
relatively narrow mass distribution~\citep{bailyn98} around
7\,M$_\odot$.  Recent analyses~\citep{ozel10,Farr11} have used the
expanded current samples of black-hole measurements in both
Roche-lobe-overflow and wind-driven X-ray binaries and proposed
distributions to fit the observations (without
quantitative consideration of selection effects).  The statistically favored models have
mass distributions that extend to high
masses ($\sim15$--$20\,M_\odot$), depending on whether the wind-fed systems with
more massive BHs (which are more uncertain) are included. Both
studies conclude that there is clear evidence for a
low-mass gap in the distribution, with no remnants found in between the maximum NS mass
($\sim2\,\mbox{M}_\odot$) and 4--5$\,\mbox{M}_\odot$. Specifically,
~\citep{Farr11} report that 
the minimum BH mass lies in the range
4.3--4.5$\msun$ at 90\% confidence level. 

By combining information from stellar evolution and core-collapse
calculations, theoretical investigations allow us to calculate the
mass of compact remnants at formation.  Timmes et al.(1996) based
their estimates on the iron core masses predicted from their stellar
evolution models.  These masses had a bimodal distribution,
causing them to predict a bimodal distribution of remnant masses.
Although they mention that fallback would be expected to broaden the
distribution, many observational analyses arguing for narrow 
distributions were nonetheless affected by their reported bimodality.  Two of the
assumptions in Timmes et al.(1996) lead to biases in their
theoretical mass estimates of compact remnants.  First, the sharp
change in iron core masses is not as pronounced in more modern stellar
models (e.g. Young et al. 2005).  Second, the final remnant mass is
determined by the entropy profile (and hence density/temperature
profiles) in the core.  Although the entropy in the core is roughly
correlated with the iron core mass, there does not exist a one-to-one
correspondence.  Using the iron core mass to estimate remnant masses
can lead to erroneous results.

Fryer \& Kalogera (2001) estimated remnant masses using collapse
calculations to guide the relation between initial stellar mass and
final remnant mass.  These estimates predicted a broader range of
NS and BH masses: for NSs, we found a
strong, dominant peak at 1.3--1.4$\msun$, with a
significant tail out to the maximum neutron-star mass. This prediction
has subsequently been borne out for NSs. For BHs they
found an extended, continuous exponential distribution without a mass gap. The
mass gap could be introduced only if a discontinuity exists in the relationship
between the supernova
explosion energy and progenitor mass. The bulk of the analysis by
Fryer \& Kalogera (2001) focused on BHs from stars without
mass loss.  Especially at higher metallicities, the final BH
mass distribution depends sensitively on the mass loss from stellar
winds~\citep{Fry01,Bel10a}.  In this paper we take advantage of new
results in both stellar evolution and stellar explosions to produce
more detailed estimates of compact remnant distributions.


To construct remnant mass distributions, we consider a wide range of
explosion mechanisms based on our current understanding of supernova
and gamma-ray burst explosions.  For an initial explosion we adopt
the convection-enhanced neutrino-driven paradigm, studying the range
of results within this mechanism.  Specifically, we study two
extremes: {\em fast-convection} explosions where explosions only occur if they
happen in the first 250\,ms after bounce, and {\em delayed-convection}
explosions which can occur over a much longer timescale (e.g., explosions
dominated by the standing accretion shock 
instability, or SASI).  In both cases, by assuming the energy input is limited
to the convective region, this model places limits on the supernova
explosion energy of up to a few times $10^{51}$\,erg.  

On top of this basic mechanism, we discuss two post-explosion engines:
magnetar or similar neutron-star driven outbursts and collapsar black
hole engines.  For example, a variant of the basic engine assumes
that additional energy is released (e.g., from energy stored in the oscillating
or rotating proto-NSs) after the launch of the convection-enhanced explosion.
This variant will have minimal
fallback, and can produce explosions well beyond the few times
$10^{51}$\,erg limit for our basic model.  Finally, we expect that some
BH forming systems will develop jets, producing long-duration
GRBs and ejecting stellar material, limiting the final mass of the
BH.  In what follows, we present compact mass distributions
from our basic engines plus a number of post-explosion variants.

In this paper we present a review of the basics of the supernova engine
(section~\ref{sec:supernovaengine}), followed by a derivation and
discussion of the energies produced by these engines (section~\ref{sec:energy}).
We derive remnant
masses for single stars as a function of progenitor mass, ranging from
the lowest mass, electron capture supernovae, to massive stars
(section~\ref{sec:reminit}).  With this remnant mass relation, we
derive remnant mass distributions for single stars
(section~\ref{sec:reminit}).  Since the observed mass distribution is
only observed in binary systems (binary pulsars and X-ray binaries), we also
derive fits to our mass 
derivation calculations appropriate for binary calculations
(section~\ref{sec:binary}).
We conclude by comparing our different models of remnant mass distributions with 
observations.

\section{Supernova Engines}
\label{sec:supernovaengine}

To understand the remnant mass distribution we must first 
understand the physical scenarios behind their formation.  And to understand 
the formation of remnants from stellar collapse, we must understand 
the explosions from stellar collapse.

Most current studies of core-collapse supernovae have focused on a
neutrino-driven engine enhanced by convection above the proto-neutron
star surface~\citep{Her94,FW04,BJ06,Bur06,FY07,Sch08,Bru09}.  It was this
type of physical engine that served as the basis for the mass distributions used
already by \cite{Fry01}.  However, this basic supernova engine
has been refined in the past decade, and it is important to
incorporate these refinements into our remnant mass distribution.

The remnant formation process can be split into 3 phases: stellar
collapse and bounce, convective engine, and post-explosion fallback.
The collapse occurs when the stellar core begins to compress under its
own weight.  The compression ultimately leads to electron capture
(which removes the degeneracy pressure support of the core) and
dissociation of the core elements into alpha particles (which removes
the thermal support).  These processes accelerate the compression which
then accelerates the rate of electron capture and iron dissociation,
leading to a runaway collapse with velocities comparable to the speed
of light.  The collapse halts when the core reaches nuclear densities
and nuclear forces (along with neutron degeneracy pressure)
dramatically increase the pressure.  This abrupt halt causes a bounce
shock to move out of the core, starting at $\sim
0.9\pm0.2$\,M$_\odot$\footnote{Note that unless specifically stated
otherwise, all masses given here are baryonic masses. The gravitational 
mass is $\sim$10-20\% lower.}  The shock
moves out until neutrino losses (and dissociation of material hit by
the shock) sap its energy reservoir, causing it to stall (at roughly
$1.1\pm0.2$\,M$_\odot$).  If the shock can be quickly revived (likely
for stars below 11\,M$_\odot$), then remnant masses would be close to
$1.1\pm0.2$\,M$_\odot$. 

For stars with initial masses above 11\,M$_\odot$, a number of
instabilities can develop in the region between the proto-NS
where the shock is launched and the position where it stalls: e.g.,
the Rayleigh-Taylor and the standing accretion shock instability
(SASI). These instabilities can convert the energy leaking out of the
proto-NS in the form of neutrinos to kinetic energy pushing
the convective region outward.  A supernova explosion occurs if the
energy in this convective region can overcome the ram pressure of the
infalling stellar material.  This is what we mean by a
``convection-enhanced, neutrino-driven supernova explosion''.  In this
paper, our basic model assumes that the energy in the supernova
explosion is the energy stored in this convective region.

The time when the energy in the convective region overcomes the ram
pressure determines the amount of material accreted onto the proto-neutron
star during the convective phase.  The total energy in the convective
region when this occurs determines the amount of energy in the explosion,
and ultimately the amount of post-shock-launch fallback.  As the
material moves outward, it pushes onto the material above it and causes
that material to accelerate.  The work done by this shocked material
slows it down, and some of the material is decelerated below the local
escape velocity.  This material will then fall back onto the
proto-NS, adding to the remnant mass.  In general, the
majority of this fallback occurs within the first 20\,s after the
explosion (Zhang et al. 2008, Fryer 2009).  The mass amount (and to
some extent the timing) of this fallback depends on the explosion
energy and the structure of the star.  It is this fallback that is the
dominant cause for the broad range of NS and BH
masses.  Determining the amount of fallback requires understanding 
the explosion itself.

\subsection{Explosion Mechanisms and Supernova Energies}
\label{sec:energy}

To determine the energy of our convection-enhanced, neutrino-driven
engines, we assume that the explosion energy is equal to the energy
stored in the convective region at the time of collapse.  Fryer (2006)
has already estimated this energy and we review this derivation here.
Colgate et al. (1993) found that they could approximate the convective
region as a roughly constant entropy atmosphere bounded by the
proto-NS and the shock region of the infalling stellar envelope.  The
structure of this atmosphere is then well defined, with the pressure
given by:
\begin{equation}
P(r) = [0.25 M_{\rm NS} G (S_{\rm rad}/S_0)^{-1}(1/r - 1/r_{\rm
    shock}) + P^{1/4}_{\rm shock}]^4 {\rm erg \, cm^{-3}}
\end{equation}
where $M_{\rm NS}$ is the proto-NS mass, $G$ is the
gravitational constant, $S_{\rm rad}$ is the entropy in Boltzmann's
constant per nucleon, $S_0 = 1.5\times10^{-11} {\rm k_B \, per \, nucleon}$, 
and $r_{\rm shock}$ and $P_{\rm shock}$ are the radius and pressure of the
accretion shock forming the outer bound of the convective region.  
$P_{\rm shock}$ is set to the ram pressure of the infalling material.  
If we assume the infalling star is accreting at the free-fall rate, 
mass continuity gives (Fryer 2006):
\begin{equation}
P_{\rm shock}(r) = 1/2 \rho_{\rm shock} v_{\rm free-fall} = (2 G
M_{\rm NS})^{0.5} \dot{M}_{\rm acc}/ (8 \pi r^{2.5}_{\rm shock})
\end{equation}
where the free-fall velocity $v_{\rm free-fall}$ is determined by the
proto-NS mass and the accretion rate $\dot{M}_{\rm acc}$ is
determined by the structure of the progenitor star assuming a
pressure-less collapse (which is a good approximation for the infall).  For a
radiation-dominated gas, the internal energy density is $3 \times
P(r)$ (Fryer 2006):
\begin{eqnarray}
u_{\rm convection}(r) &=& 3 \left[4.7 \times 10^8 \frac{M_{\rm
      NS}}{M_\odot} \frac{10 k_{\rm B} {\rm nucleon}^{-1}}{S_{\rm rad}}
  \left( \frac{10^6 {\rm cm}}{r} - \frac{10^6 {\rm cm}}{r_{\rm shock}}
  \right) + \right. \nonumber \\
  && \left. 1.2 \times 10^6 \left( \frac{M_{\rm NS}}{M_\odot}
  \frac{\dot{M}_{\rm acc}}{M_\odot \rm{s}^{-1}} \right)^{1/4} \left(
  \frac{2\times 10^7 {\rm cm}}{r_{\rm shock}} \right)^{5/8}\right]^4
{\rm erg \, cm^{-3}}.
\end{eqnarray}
Integrating over the entire atmosphere, we can derive the maximum
energy stored in the convective region.  If the energy in the
convective region is above this maximum value, the pressure in the
atmosphere will be larger than the infall pressure, causing the shock
radius to expand.  By the time the shock expands to 1000\,km, the
expansion is really an explosion.  After the explosion is launched,
the atmosphere is too thin to absorb much neutrino energy.  Neutrinos 
leaking from the core can no longer add much energy to the now exploding 
atmosphere.  The energy of the explosion from the neutrino-driven mechanism 
is then limited to roughly the energy when the explosion is launched.

\begin{figure}[htbp]
\plotone{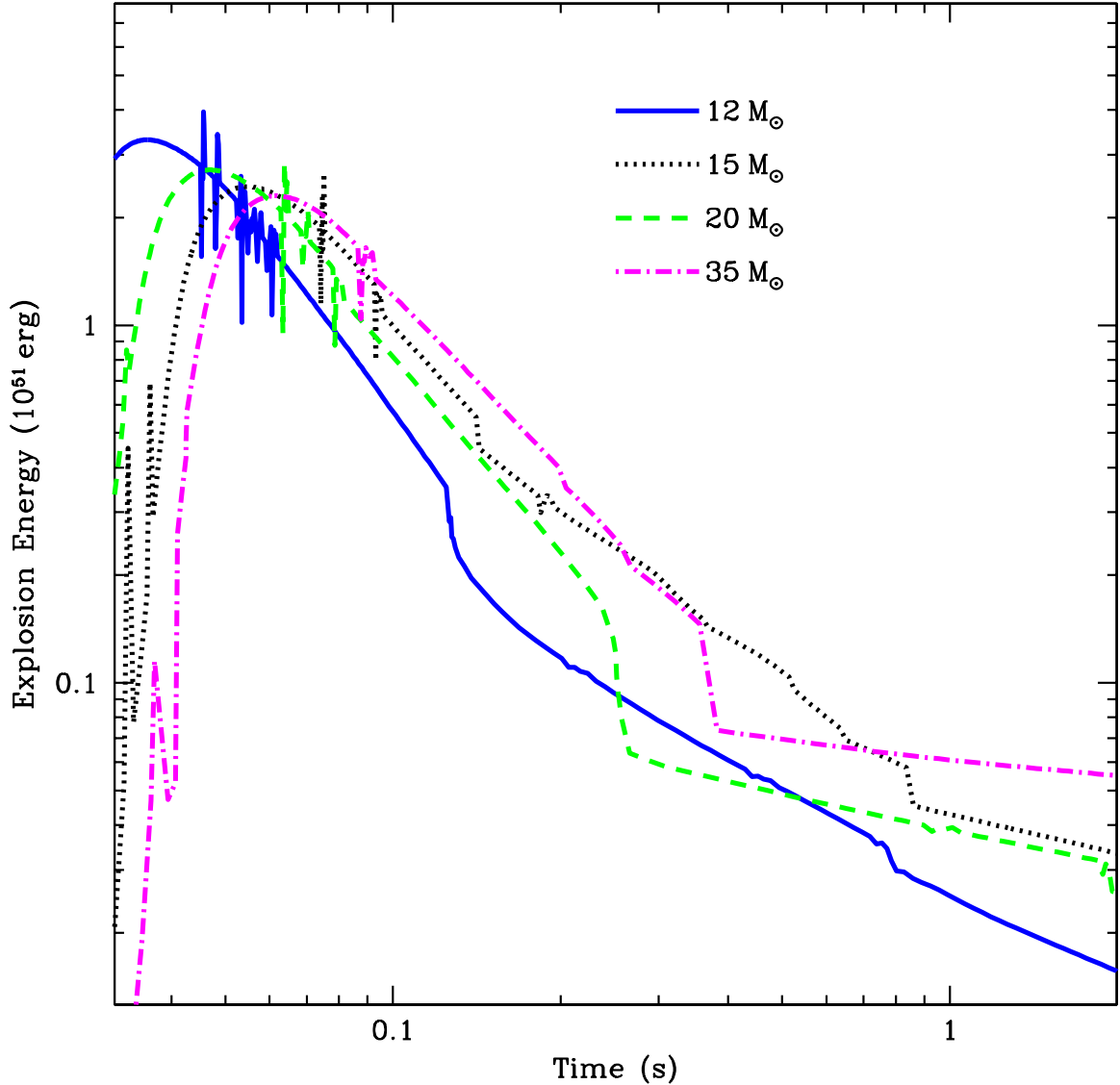}
  \caption{Energy stored in the convection region as a function of
    time after bounce.  For the convection-enhanced neutrino
    mechanism, this energy is the available energy to power a
    supernova explosion.  With a peak near $\sim 3-4 \times
    10^{51}\,{\rm erg}$, this energy provides a natural explanation
    for why ``typical'' supernova have energies of roughly
    $10^{51}$\,erg even though the total potential energy released is
    closer to $10^{53}$\,erg.  Note, however, that it is difficult to
    make a strong explosion after a long delay.}
  \label{fig:snevst}
\end{figure}

For a typical atmospheric entropy of $10 \, k_{\rm B} \, {\rm
  nucleon}^{-1}$ and shock radius of 1000\,km, energies above
$5\times10^{50}$\,erg only occur for accretion rates above 1\,$M_\odot
{\rm s^{-1}}$.  See Fryer (2006) for a study of the dependence of the
explosion energies on the atmosphere parameters.  The energy that
can be stored in the convective region decreases as the accretion rate
decreases.  The accretion rate of the infalling stellar material
decreases with time, causing the total available explosion energy to
decrease with time (Fig.~\ref{fig:snevst}).  If the delay in the explosion 
is long, the explosion will be weak.

This model assumes the explosion energy is stored in the convective
region and provides a natural explanation for why the observed explosion
energy for most supernovae are found in the range $\sim0.5$--$2 \times
10^{51}\,{\rm erg}$ even though the potential energy released in
stellar collapse is $\sim 10^{53}$\,erg.  The peak energy stored in
the convective region is roughly a few times $10^{51}$\,erg, and this
engine cannot produce a stronger explosion.  If the
explosion occurs less than 250\,ms after bounce, the energy is above
$10^{51}$\,erg for most progenitors.  Most observed supernovae have explosion
energies within these limits.  Magnetohydrodynamic engines produce 
a much broader range of supernova engines and are thus not the likely 
engines behind ``standard'' core-collapse supernovae.

Our basic models consider only the convection-enhanced
neutrino-driven supernova engine, since this is able to account for the
majority of supernovae. Within this basic set, we still vary
the length of the delay time prior to the explosion.  One model
allows for considerable contribution from the SASI engine, a
current focus of many supernova groups (Blondin et al. 2003, Burrows
et al.  2007, Bruenn et al. 2009, Scheck et al. 2008, Marek \& Janka
2009).  The SASI has a growth time that is typically over an order of
magnitude longer than the Rayleigh-Taylor instabilities, and hence is
unable to drive an explosion at early times.  For our standard model,
we allow an explosion only after the accretion rate drops below
1\,$M_\odot {\rm \, s^{-1}}$. In this formalism, stars below about
$15\,M_\odot$ produce explosions enhanced by convection dominated by
the Rayleigh-Taylor instability, whereas above this mass the explosions
are enhanced by convection dominated by large-scale modes
characterized by the SASI.  For many progenitors, the latter occurs at later
times.  If the proto-NS exceeds 3\,$M_\odot$, we assume the
star collapses promptly to a BH (a.k.a. the ``direct'' BH
formation scenario from Fryer \& Kalogera 2001).

Our alternate convection-enhanced supernova engine model focuses on
strong convective models developing rapidly and driving an explosion
within the first 250\,ms (e.g., Herant et al 1994, Fryer \& Warren
2002).  In this engine, we assume an explosion occurs when the
accretion rate drops below 3\,$M_\odot {\rm s^{-1}}$.  If the
proto-NS exceeds 3\,$M_\odot$ or the delay time exceeds
250\,ms, we assume the star collapses to a BH.

\begin{figure}[htbp]
\plotone{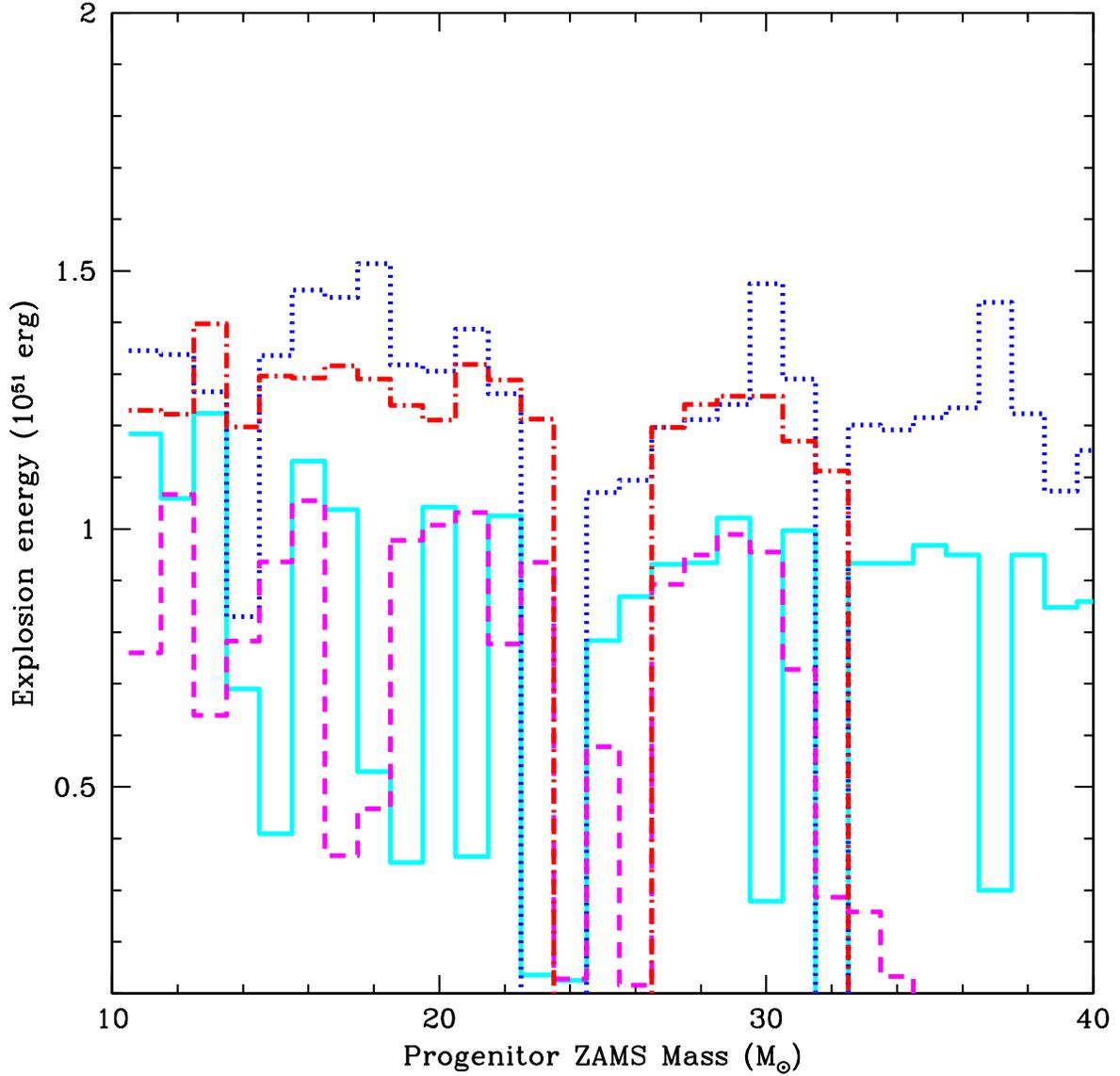}
  \caption{Predicted explosion energy as a function of progenitor mass
    for two engine models and two progenitor metallicities: solar
    metallicty progenitor with delayed explosion (solid), solar
    metallicity progenitor with rapid explosion (dotted), zero
    metallicity with delayed explosion (dashed), zero metallicity with
    rapid explosion (dot-dashed).  The rapid explosions are more
    energetic when they succeed, but are more likely to fail completely.}
  \label{fig:snenergy}
\end{figure}

Figure~\ref{fig:snenergy} shows the supernova explosion energies for
these two models as a function of progenitor mass using the Woosley et
al.(2002) progenitors at solar and zero metallicity.  The delayed
explosions tend to be weaker than our rapid explosion model, producing
many more explosions with energies below $10^{51}$\,erg.  The
explosion energies depend upon the prescriptions for mass loss (or,
more accurately, the coefficients used in the mass loss prescription)
and convection, and we discuss the differences in these models in
section~\ref{sec:caveats}.  Above $30$--$35\,M_\odot$, the mass loss 
plays the deciding role in determining the remnant mass, as
we discuss this in detail in section~\ref{sec:30msun}.

After the initial launch of the convection-enhanced explosion, the
region above the proto-NS is evacuated.  At this point, the
energy deposited by neutrinos in this region is drastically reduced.  
As we discuss below (section~\ref{sec:energy}),  this places a limit
on the explosion energy from the neutrino driven supernova engine 

There exist other mechanisms to extract the enormous energy reserved
in the collapsed stellar core.  For example, if the stellar core is
rotating rapidly, strong magnetic fields may develop.  As the
proto-NS contracts, an enormous amount of energy can be stored in the rotation,
driving a second outburst from the NS. 
This energy source has been invoked to explain a variety of energetic
supernova explosions (e.g. Maeda et al. 2007, Kasen \& Bildsten 2010).
Although quantitative calculations of rotation powered explosions do not exist,
we can place an upper limit based on the predicted spin-energy in
the NS ($\sim 3\times10^{52}$\, erg for a millisecond
NS). If such an explosion occurs, the fallback onto the
proto-NS will be minimal.  In our third (``Magnetar'') model 
for remnant distributions, we assume this additional explosive engine
occurs in all systems.

We note that fallback can also cause a secondary outburst as
the material accretes onto the protoNS~\citep{Fry06}.  This
is less explosive than a magnetar and, to show the extremes, we focus
only on the magnetar explosion.

The collapsar engine behind hypernovae posits that a rapidly spinning
star collapses down to a BH.  The jet produced by the
accretion disk that forms around the BH ``drills'' through the
stellar envelope and ultimately ejects the star, preventing further
accretion (MacFadyen \& Woosley 1999).  In our final scenario
(Collapsar), we include the mass loss from this powerful engine.

There exist many permutations of our basic convection engines combined with
post-explosion engines (e.g. collapsar, magnetar).  However, it is
anticipated that just a small fraction of systems are affected by
the post-explosion engines.  To encapsulate the effect of the post-explosion
models on the final mass distributions, we assume (an unreasonably high) 100\% efficiency in
these engines, and study only the four cases described shown in Table~\ref{table:engines}.

\begin{deluxetable}{lcccc}
\tablewidth{0pt}
\tablecaption{Core Collapse Engines}
\tablehead{
  \colhead{Supernova}
& \colhead{No Additional}
& \colhead{Collapsar}
& \colhead{Fallback}
& \colhead{Magnetar} \\

\colhead{Engine}
& \colhead{Explosion}
& \colhead{Engine}
& \colhead{Outburst}
& \colhead{Engine}
}
\startdata
Rapid & {\bf Rapid} & {\bf Rapid w/} & & \\ 
      &             & {\bf Collapsar} & & \\ 
Delayed & {\bf Delayed} & & & {\bf Delayed w/} \\
        & {\bf Delayed} & & & {\bf Magnetar} \\

\enddata
\label{table:engines}

\end{deluxetable}

\section{Remnant Masses from Single-Star Collapse}
\label{sec:reminit}

We are interested in determining the masses of the compact remnants
(NS and BH). Doing this requires a combined understanding of stellar
structure, the supernova engine, and the propagation of the explosion
through the star (along with the fallback material).  These are all
very active areas of research, and in this section we use the current
understanding of these physical processes to derive remnant mass
distributions.  The physics determining the remnant masses can be
divided into three zero-age main sequence (ZAMS) progenitor mass
ranges: stars below 11\,M$_\odot$, stars between 11 and
$\sim$30\,M$_\odot$, and star more massive than $\sim$30$--$35
\,M$_\odot$.  For the low mass stars, the primary uncertainty in the
remnant mass arises from uncertainties in stellar convection.  For the
intermediate masses, our understanding, or lack thereof, of the
supernova explosion mechanism dominates the uncertainties.  For the
massive stars the primary uncertainty is in the prescription for mass
loss (for details, see section~\ref{sec:caveats}).  To derive remnant
masses we are forced to make a range of assumptions (described in
detail below). We note that understanding the nature of these
uncertainties is a critical component in using observational masses to
constrain theory.

\subsection{Stars Below $\sim$ 11\,M$_\odot$}

For ZAMS progenitor stars below $\sim$11\,M$_\odot$, the fate of
collapse is the same for all of our explosion models.  The fallback in
the supernova explosion from these progenitors is negligible.  As
such, the uncertainties in the shock propagation are minimal.  In
addition, as we shall see below, the uncertainties in the explosion
mechanism only alter the final remnant mass by, at most,
0.1\,M$_\odot$.  The primary uncertainty in the remnant mass
distribution from these stars arises from determining the number of
these objects which, in turn, is caused by uncertainties in stellar
evolution.  In particular, stellar evolution models do not accurately
constrain the transition mass between white dwarf and NS formation
(i.e., the lower mass limit for NS formation).

The exact value of the lower limit ($M_{\rm NS}^{\rm lower}$) for
core-collapse has been a matter of debate for many decades (see Iben
\& Renzini 1983 for a review).  The metallicity dependence for this
limit is even more controversial.  Heger et al. (2003) argued that the
metallicity dependence was negligible, and that the lower limit was
roughly at 9\,M$_\odot$ for all metallicities.  Poelarends et
al. (2008) have studied this matter in more detail, and have a range
of solutions for the metallicity dependence.\footnote{At the dividing
  line between ONe white dwarf formation and iron core-collapse
  supernovae lies a class of ``electron-capture'' supernovae in a
  narrow mass range ($\sim$0.2$--$0.3\,M$_\odot$---see Poelerands et
  al. 2008).  Electron-capture supernovae occur in ONe cores when
  electron capture triggers a core-collapse prior to neon ignition.
  The electron capture supernova and remnant are similar to those in
  the low-mass iron core-collapse supernova produced by progenitors
  below 11\,M$_\odot$. As such, for the purpose of our remnant
  calculation we will lump this class of supernova with all other
  supernovae below 11\,M$_\odot$.}  Their ``preferred'' model predicts
that $M_{\rm NS}^{\rm lower}$ drops from 9\,M$_\odot$ at solar
metallicity down to below 6.3\,M$_\odot$ at $Z_{\rm
  metal}=10^{-3}Z_{\odot}$.   We use the
following fit to their solution:
\begin{eqnarray}
M_{\rm NS}^{\rm lower} &=& 9.0 + 0.9 {\rm log_{10}} (Z_{\rm
    metal}/Z_{\odot}) M_\odot \; {\rm if} \, {\rm log_{10}} (Z_{\rm
    metal}/Z_{\odot}) > -3 \nonumber\\ && 6.3 M_\odot \; {\rm
    otherwise}.
\end{eqnarray}
The fraction of NS/BH remnants (assuming a Salpeter initial mass
function) produced by stars below 11\,M$_\odot$, as a function of
metallicity, is shown in Figure 3.  However, we note that Poelerands et
al. (2008) found that this result was highly sensitive to their
prescriptions for mass loss and dredge up.  In some of their models, 
the variation with respect to metallicity less than 1\,M$_\odot$ 
in contrast to the 2.7\,M$_\odot$ assumed in this model.

Understanding stellar evolution will allow us to determine how many
NSs are formed from stars with initial masses below 11\,M$_\odot$.
However, the mass of the remnant that is formed in the collapse of
these stars is determined by the supernova engine.  For these
low mass stars (both electron capture/ONe supernovae and iron collapse
supernovae) the envelope of the star is fairly tenuous (and relatively
easily ejected), and the collapsing core is close to the Chandrasekhar
limit: $\sim 1.38$\,M$_\odot$ for OMgNe stars.  The remnant mass is
the Chandrasekhar mass minus any ejecta mass.  Simulations of the collapse
of these objects have found that electron capture supernovae can eject
from 0.01--0.2\,M$_\odot$ of stellar material (see Fryer et al. 1999 for
a review, or more recent work by Kitaura et al. 2006, Dessart et
al. 2007).  The favored model by Fryer et al. (1999) predicted an
ejecta mass of roughly 0.1\,M$_\odot$, leading to a baryonic mass of the
collapsed NS of $\sim$1.28\,M$_\odot$.  The recent results by Kitaura
et al. (2006) suggest that the ejecta is closer to 0.02\,M$_\odot$,
predicting a final remnant mass of $\sim$1.36\,M$_\odot$.  Low-mass 
iron core-collapses are likely to be similar in mass.  Although we
adopt the former as our default mass distribution, we discuss both in
our mass distribution calculations.  We expect the remnant mass to be 
relatively flat for all stars below 11\,M$_\odot$. 

\subsection{Stars between $\sim 11$\,M$_\odot$ and  $\sim30$\,M$_\odot$}

The remnant masses of stars more massive than $\sim$ 11\,M$_\odot$ are
primarily determined by the amount of material that falls back onto
the proto-NS after the launch of the explosion.  As the ejecta pushes
out against the surrounding stellar envelope it loses energy, and some
of this material falls back onto the
compact remnant.  The previous piston-driven explosions both delayed
and underestimated this fallback for a given supernova explosion
(Young \& Fryer 2007).  Energy-driven explosions more accurately mimic the likely
convection-enhanced explosion mechanism, and lead
to a solution where most of this fallback occurs in the first
10--20\,s (see Fryer 2009 for a review).  For a given energy drive, the
amount of fallback can be calculated fairly accurately.

The amount of fallback depends upon the energy in the explosion.  As
mentioned before, here we study four types of engines, producing a wide
range of fallback yields.  For the initial explosion energies we use
either the delayed or rapid explosions discussed in
section~\ref{sec:energy}.  Figure~\ref{fig:enginecomp} shows the
remnant masses for these two engines using the Woosley et al. (2002)
progenitors. The broad range of energies in the delayed explosion
produces a continuous range in remnant masses, while the rapid
explosion mechanism leads to a sharp transition between neutron 
stars and more massive black holes.

For rapidly-spinning progenitors, the initial fallback can be altered by
a post-explosion engine.  In the case of BH formation, the rapidly
spinning systems will drive further outflow via the collapsar engine.
This engine ultimately disrupts the entire star, typically limiting
the black hole mass (MacFadyen \& Woosley 1999) to a maximum of 5\,M$_\odot$.
Figure~\ref{fig:enginecomp} shows a mass distribution of the
rapid-explosion initial model coupled to a collapsar engine.

Alternatively, a rapidly-spinning magnetar model can also halt
fallback in all progenitors that initially form a NS.  Here we
assume that this magneto-hydrodynamic engine prevents further
fallback.  As an example, Figure~\ref{fig:enginecomp} shows a mass
distribution of the delayed-explosion initial model coupled to a
rapidly-spinning magnetar engine.

Below $\sim$25--30\,$M_\odot$, the results are fairly insensitive to
the metallicity. Figure~\ref{fig:mremwooD} shows our 
delayed model results for a range of metallicities, with our best 
fit to the remnant masses for these metallicities:
\begin{equation}
M_{\rm remnant,delay} = 1.1 + 0.2 e^{(M_{\rm star}-11.0)/4.} - (2.0+Z_{\rm
  metal}) e^{0.4(M_{\rm star}-26.0)},
\label{eq:normald}
\end{equation}
where $M_{\rm star}$ is the initial mass of the star and $Z_{\rm
metal}$ is the metallicity with respect to solar.
The corresponding fit for our rapid explosion is (Fig.~\ref{fig:mremwooR}):
\begin{eqnarray}
M_{\rm remnant,rapid} &=& 1.1 + 0.2 e^{(M_{\rm star}-11.0)/7.5} + 10.0
(1.0+Z_{\rm metal}) e^{-(M_{\rm star}-23.5)^2/(1.0+Z_{\rm metal})^2} \nonumber\\ 
&& \;\;\;\;
{\rm if} \, M_{\rm star} < 22M_\odot \nonumber\\ &=& M_{\rm
  remnant,delay} - 1.85 + 0.25 Z_{\rm metal} + 10.0
(1.0+Z_{\rm metal}) e^{-(M_{\rm star}-23.5)^2/(1.0+Z_{\rm metal})^2} \nonumber \\
&& \;\;\;\; {\rm otherwise}.
\label{eq:normalr}
\end{eqnarray}

\subsection{Stars above $\sim$30\,M$_\odot$}
\label{sec:30msun}

Above 30\,$M_\odot$, the treatment of mass-loss from winds and the
wind metallicity-dependence become the dominant uncertainty, and
differences between the different explosion models fade.
Figure~\ref{fig:enginecomp} shows results for the Woosley et
al. (2002) models at both solar and zero metallicity. These models use the
Nieuwenhuijzen \& Jager (1990) mass loss prescription ($\dot{M}
\propto L^{1.42} M^{0.16} R^{0.81}$, where $L$ is the luminosity, $M$ is the stellar mass,
and $R$ is the stellar radius) that is used by many stellar evolution
calculations.  A metallicity dependence ($\dot{M} \propto Z_{\rm
metal}^{0.5}$) from Kudritzki (1989) is added to this prescription, a common approach
in stellar evolution codes (for stars not in the Wolf-Rayet
phase). The primary differences among stellar models is the
proportionality constant in front of these quantities.  This difference translates 
into a difference in the effective metallicity used by different 
stellar theorists:  i.e. the Woosley et al. (2002) models at 1/3 solar 
are very similar to the solar metallicity Limongi \& Chieffi (2006) 
models.  For stars less than
50\,M$_\odot$, the remnant mass is fairly well fit by:
\begin{equation}
M_{\rm remnant, delay} = {\rm min}(33.35+(4.75+1.25Z_{\rm metal})
(M_{\rm star}-34),M_{\rm star} - Z_{\rm metal}^{1/2} (1.3M_{\rm star} - 18.35)),
\label{eq:massive}
\end{equation}
where $M_{\rm star}$ is the zero-age main sequence mass of the star in
solar masses.  For our rapid explosions, the remnant is essentially 
identical:
\begin{equation}
M_{\rm remnant, rapid} = M_{\rm remnant, delay} - 1.85 + Z_{\rm metal} (75-M_{\rm star})/20.
\label{eq:massiverapid}
\end{equation}
At solar metallicity, the fate of stars more
massive than 50\,M$_\odot$ depends sensitively on the prescription for
mass loss.  Stellar theorists run very few models at these high masses,
and their results vary widely.  Fitting the Woosley et al. (2002) and
Heger et al.(2003) results at solar metallicities, we get identical 
results for both our delayed and rapid explosions:
\begin{eqnarray}
M_{\rm remnant} &=& 1.8 + 0.04\times(90-M_{\rm star}) \; {\rm if} \,
 M_{\rm star} < 90M_\odot \nonumber\\ && 1.8 + {\rm log_{10}}(M_{\rm
 star}-89) \; {\rm otherwise}.
\label{eq:verymassive}
\end{eqnarray}
For lower metallicity stars, a reasonable fit is the maximum between
equation~\ref{eq:massive} and equation~\ref{eq:verymassive}.  These
equations do not include the effect of pulsational instabilities.
\cite{Heg03} argued that below a metallicity of $\sim
10^{-3}$--$10^{-4}$ solar, pulsations eject mass for progenitors lying in
the mass range 100--140\,M$_\odot$.  At this same metallicity, stars 
with masses lying between 140--260\,M$_\odot$ are disrupted entirely 
in a pair-instability supernova, leaving behind no compact remnant 
whatsoever.  It may be that pair-instability outbursts occur at even 
higher metallicity.  We do not include these pair-instability 
explosions in this analysis.

These mass distributions do not include any additional explosions 
produced by magnetars or collapsars.  Magnetar outbursts following 
a supernova explosion will prevent fallback, turning some of our 
BH remnants into NSs.  Collapsars will limit the 
amount of BH accretion, turning massive BH systems 
into low-mass systems.

\subsection{Putting it together for Single Stars}

\begin{figure}[htbp]
\plotone{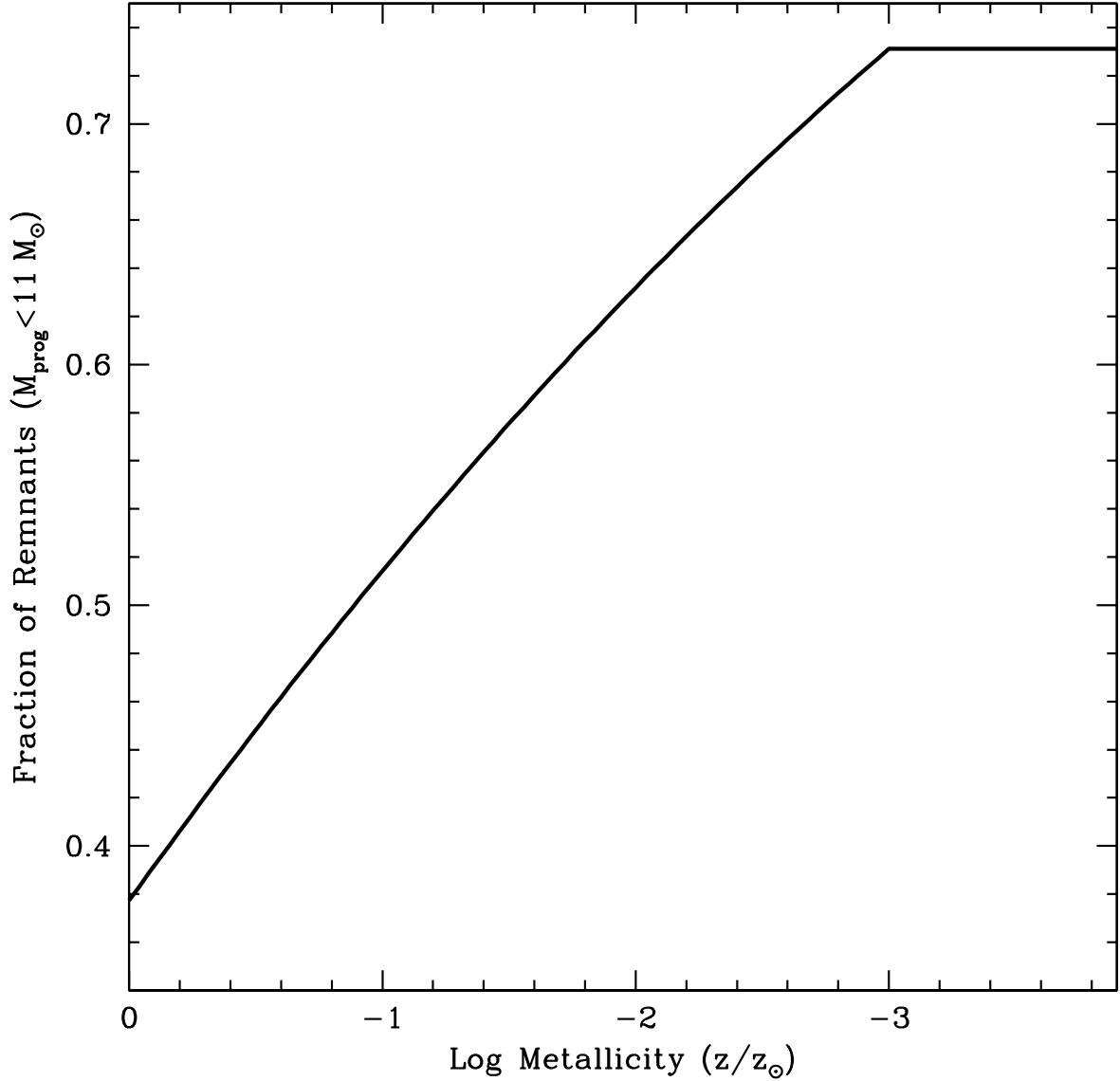}
\caption{Fraction of compact remnants produced by stars below 
11\,M$_\odot$ as a function of metallicity.  Below $\sim$0.1 solar 
metallicity, most compact remnants are produced by these stars.  
They will dominate the neutron star population.}
\label{fig:lowerlimit}
\end{figure}

\begin{figure}[htbp]
\plottwo{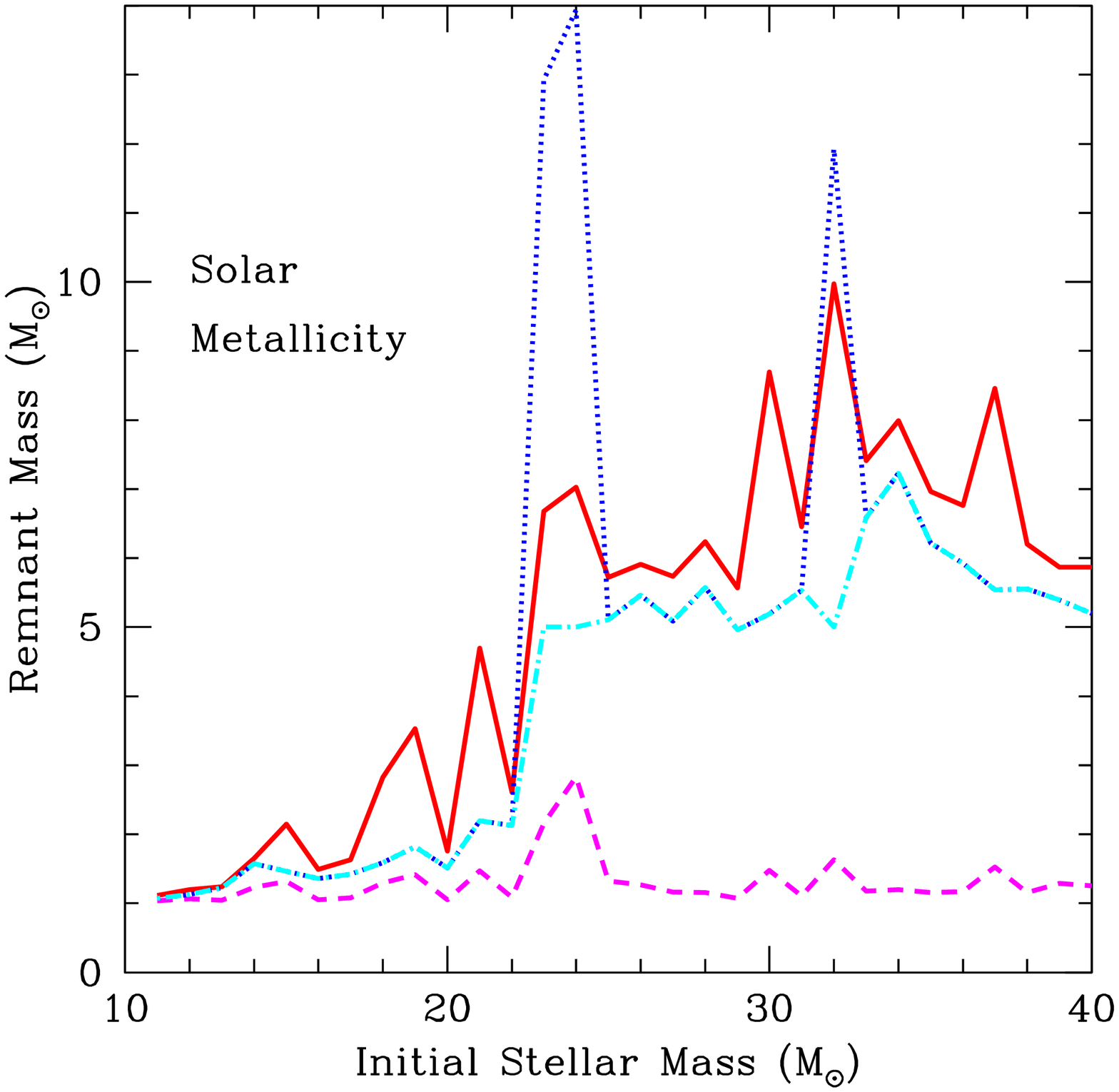}{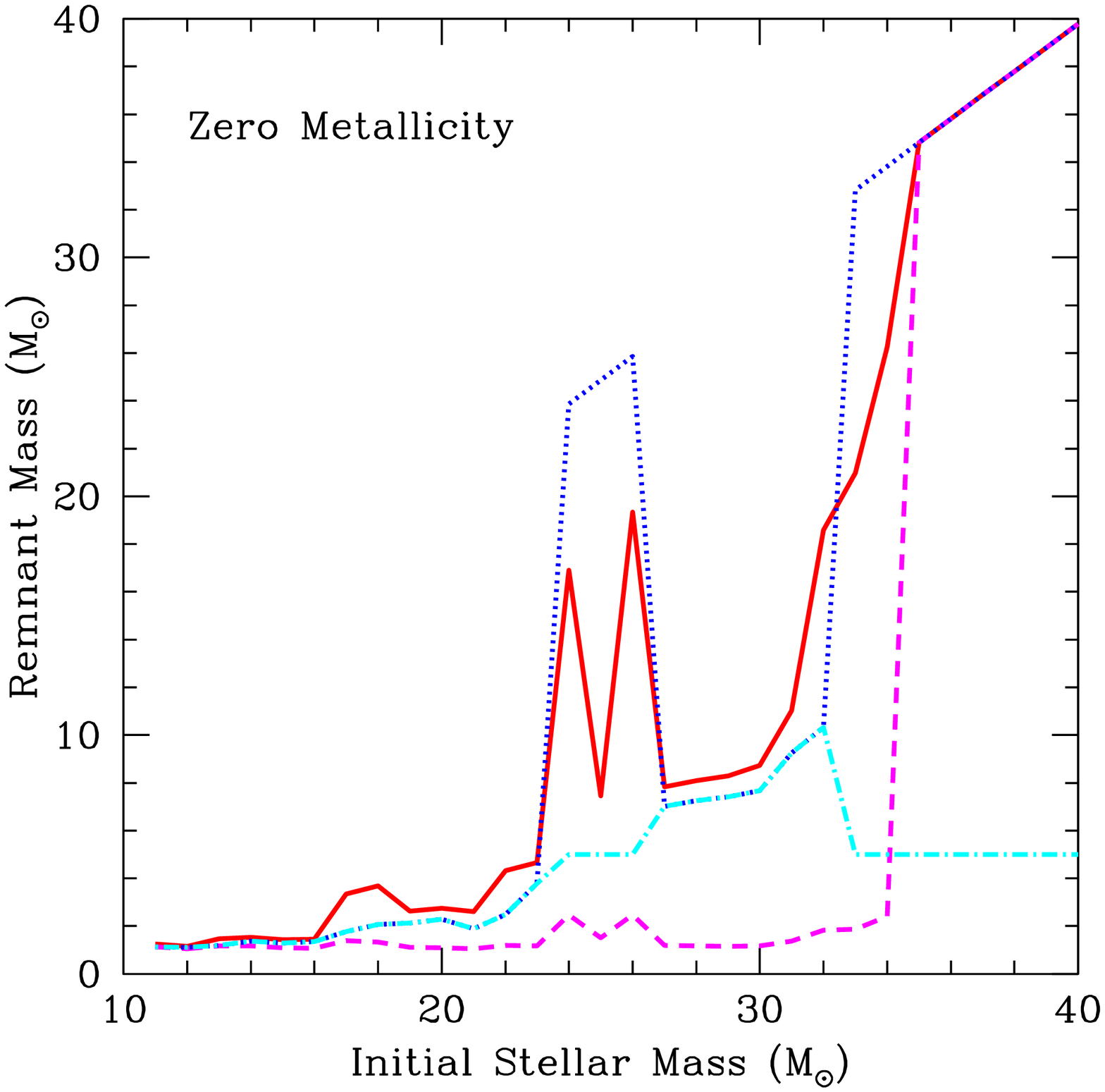}
\caption{Remnant mass versus initial stellar mass for 4 different
  explosion models: delayed explosion alone (solid), rapid explosion
  alone (dotted), delayed explosion with magnetar (dashed), rapid
  explosion with collapsar for direct BHs (dot-dashed).  The
  delayed explosion produces more intermediate mass NSs and
  low mass BHs.  The collapsar model limits the number of
  massive BHs and likely happens in only a small fraction of
  BH forming systems.  The magnetar model limits the number of
  BHs and is also likely to be rare.}
\label{fig:enginecomp}
\end{figure}

\begin{figure}[htbp]
\plottwo{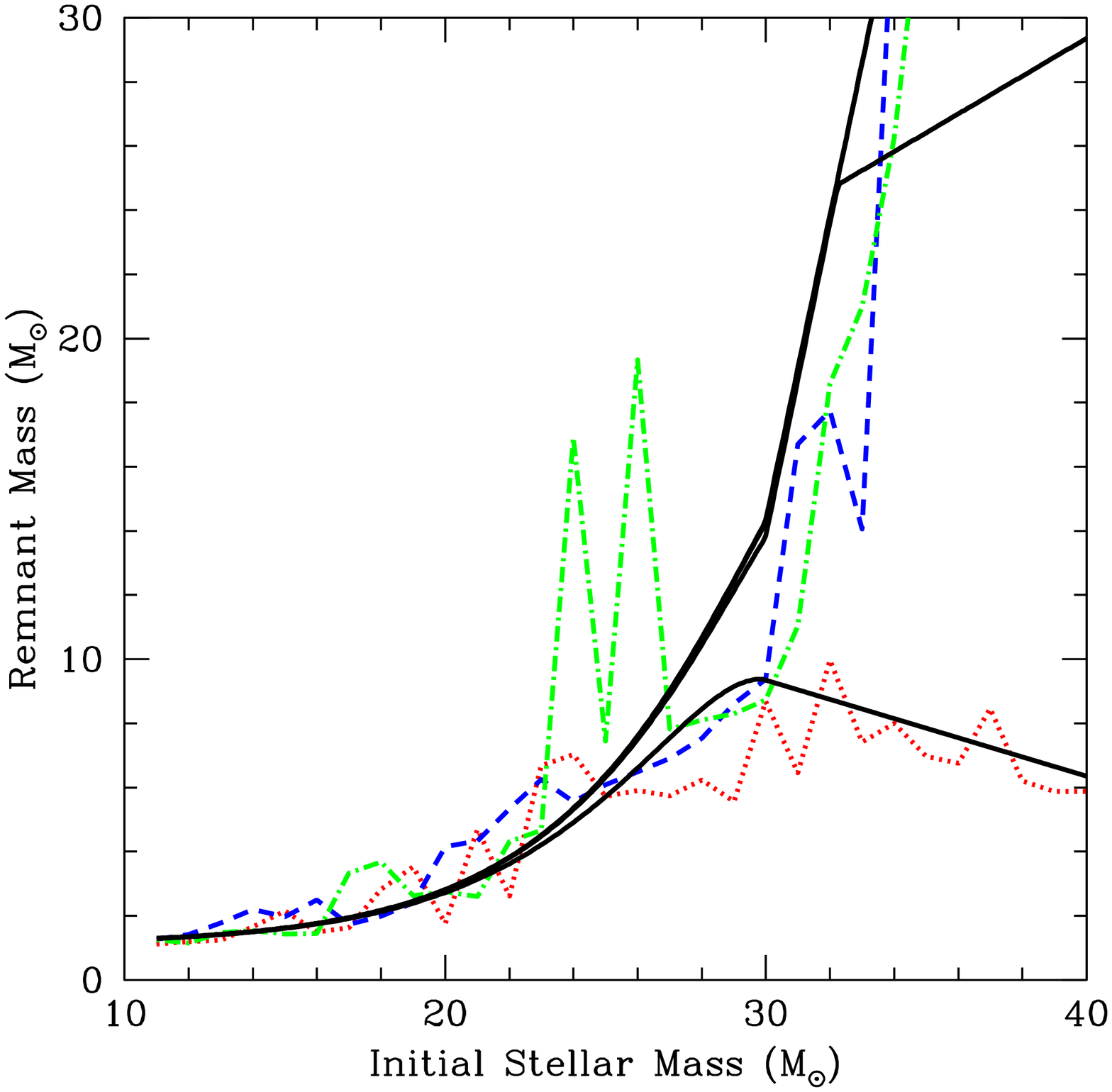}{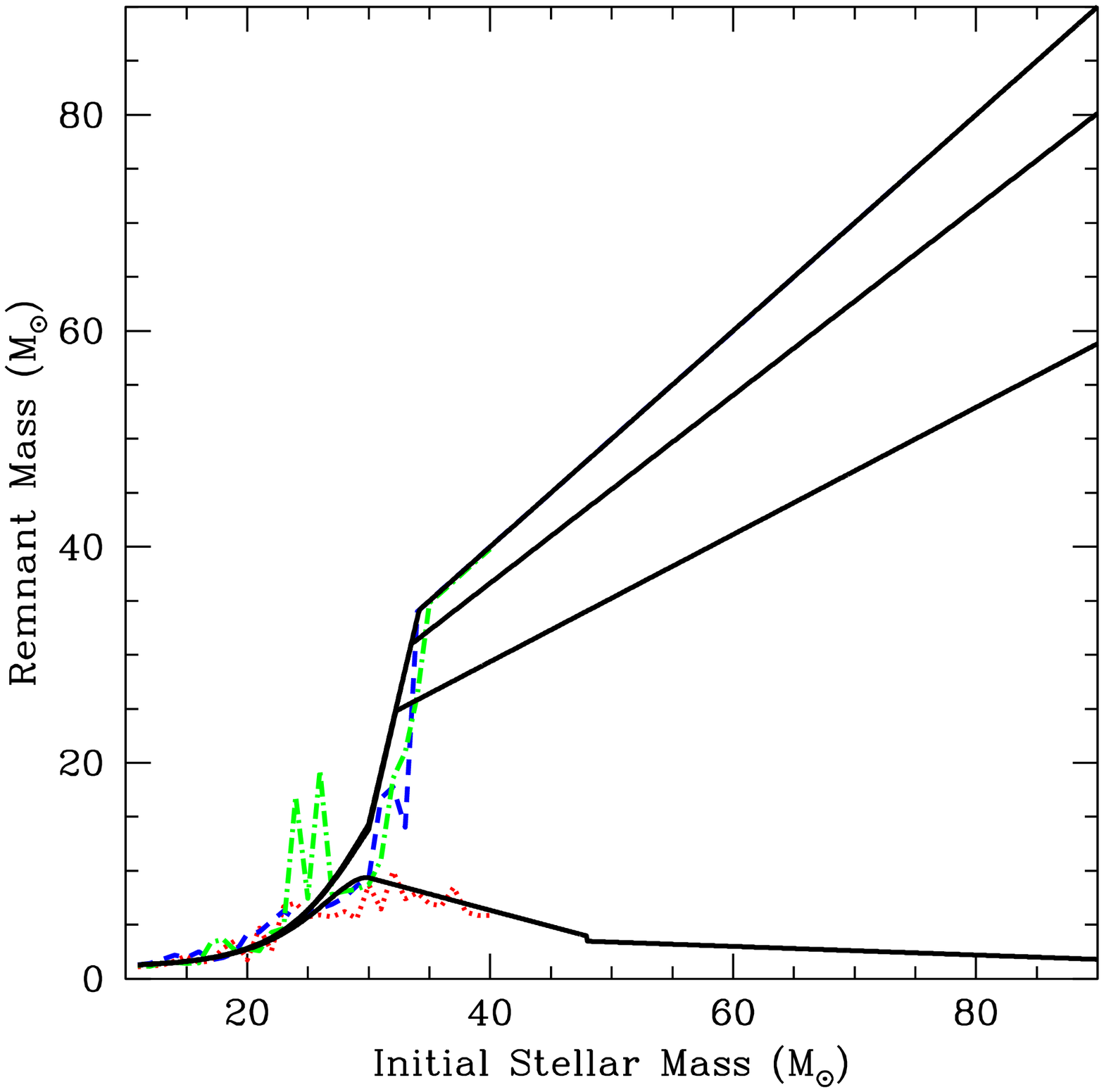}
\caption{Remnant mass versus initial stellar mass in our delayed model
  for 3 different stellar model suites from Woosley et al. (2002):
  solar metallicity (dot-dashed), $10^{-4}$ solar (dashed), and zero
  metallicity (dotted).  The solid curve shows our fit to this data
  (equations:~\ref{eq:normald},~\ref{eq:massive},~\ref{eq:verymassive}).
  Note that below 30\,M$_\odot$, metallicity has very little effect on
  the remnant mass.  Above 30\,M$_\odot$, the metallicity dependence
  of winds alters the final remnant mass.  For these high masses, the
  explosion energy is very weak, and a considerable amount of the
  hydrogen and helium envelope falls back onto the compact remnant.
  The left panel is at solar metallicity, the right panel is at zero
  metallicty.}

\label{fig:mremwooD}
\end{figure}

\begin{figure}[htbp]
\plottwo{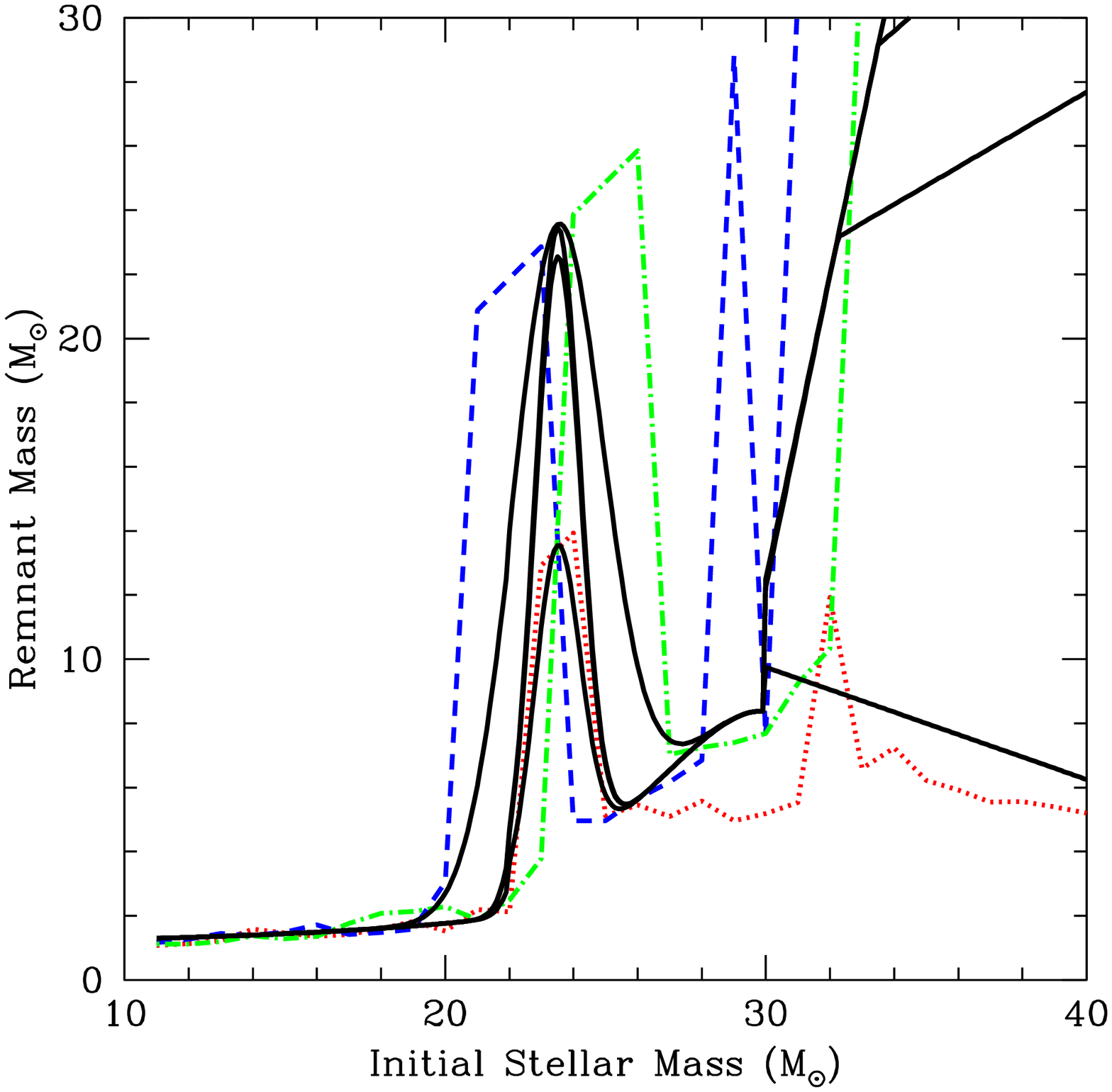}{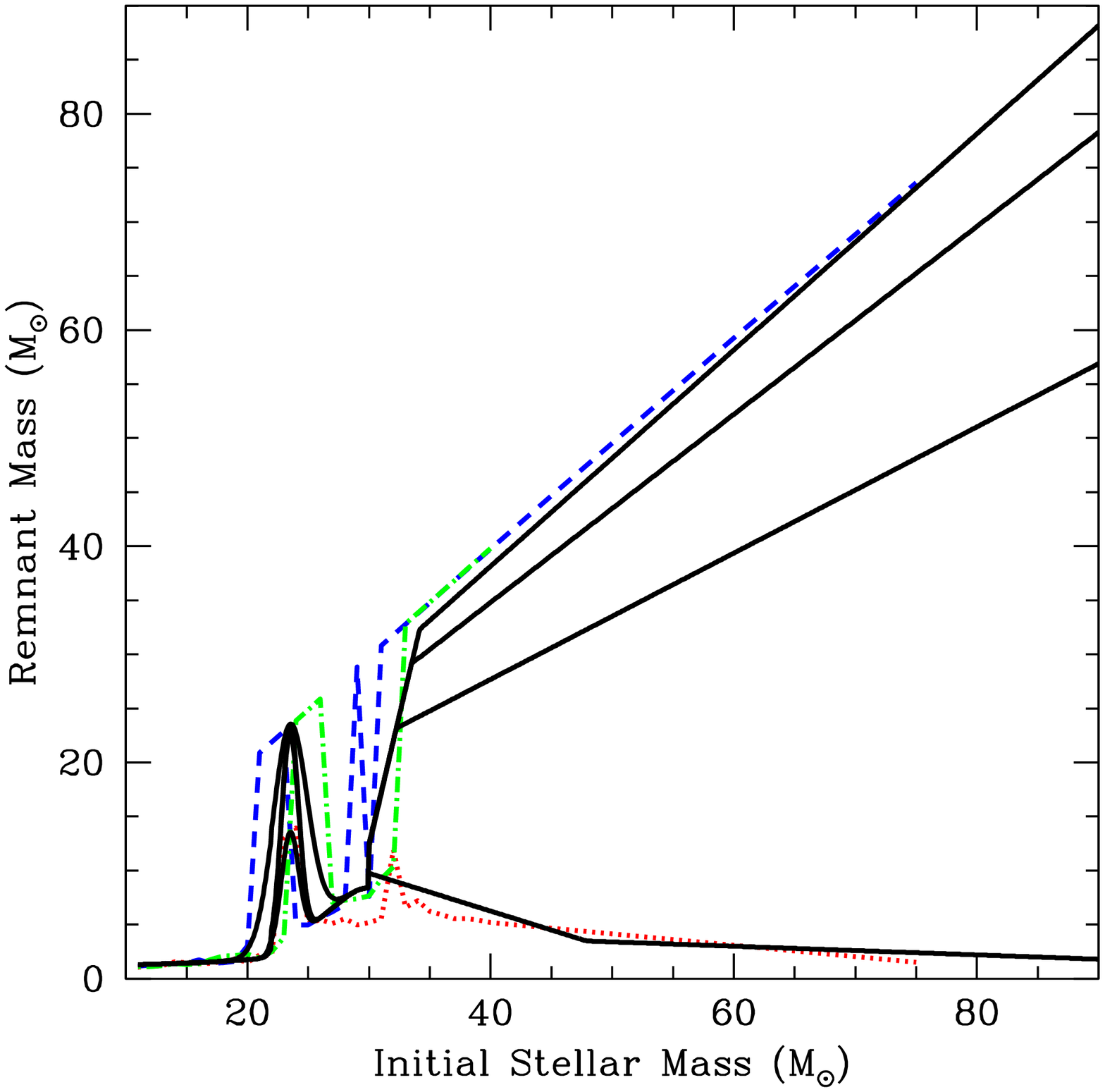}
\caption{Remnant mass versus initial stellar mass in our rapid model
  for 3 different stellar model suites from Woosley et al. (2002):
  solar metallicity (dot-dashed), $10^{-4}$ solar (dashed), and zero
  metallicity (dotted).  The solid curve shows our fit to this data
  (equations:~\ref{eq:normald},~\ref{eq:massive},~\ref{eq:verymassive}).
  Note that below 30\,M$_\odot$, metallicity only plays a role in the
  BHs formed from progenitors with masses near 23\,M$_\odot$.
  In the Woosley et al. (2002) progenitor models, the mass of these
  stars peak at solar metallicity.  More massive stars have strong
  winds that decrease the star mass and lower the core density.  Above
  25-30\,$M_\odot$, the remnant mass is very similar to that of our
  delayed model.  At high metallicities, the rapid explosions produce
  slightly less massive remnants across a wide range of initial
  progenitor mass because the stronger explosions have less fall back.  
  But at low metallicities, the remnant masses produced for both the 
  delayed and rapid explosion engines are identical.}
\label{fig:mremwooR}
\end{figure}

Our mass distribution, combining our analysis of stars above and
below 11\,M$_\odot$ (described by equations 1--4), is shown in
figures~\ref{fig:lowerlimit}, \ref{fig:mremwooD}, and
\ref{fig:mremwooR}.  Note that the primary effect of metallicity on
the NS mass distribution is to alter the number of low-mass
NSs.  This reflects the changing lower limit for neutron
star formation.  BH mass distributions, arising from more
massive stars with stronger winds, have a more dramatic dependence on
metallicity.  These masses are baryonic masses; the gravitational
masses are likely to be $\sim 10\%$ lower (e.g. Strobel \& Weigel
2001).

\begin{figure}
\plottwo{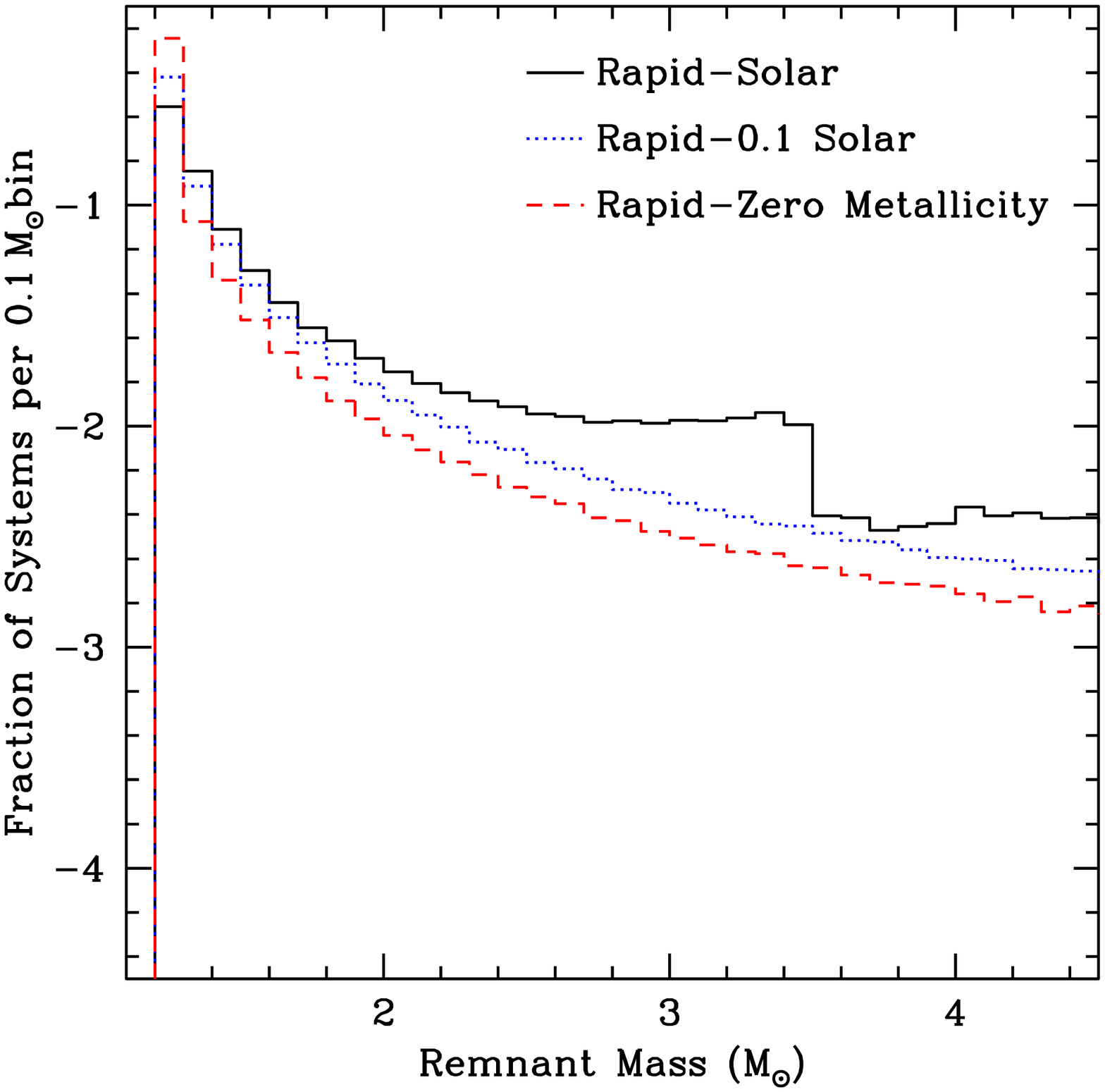}{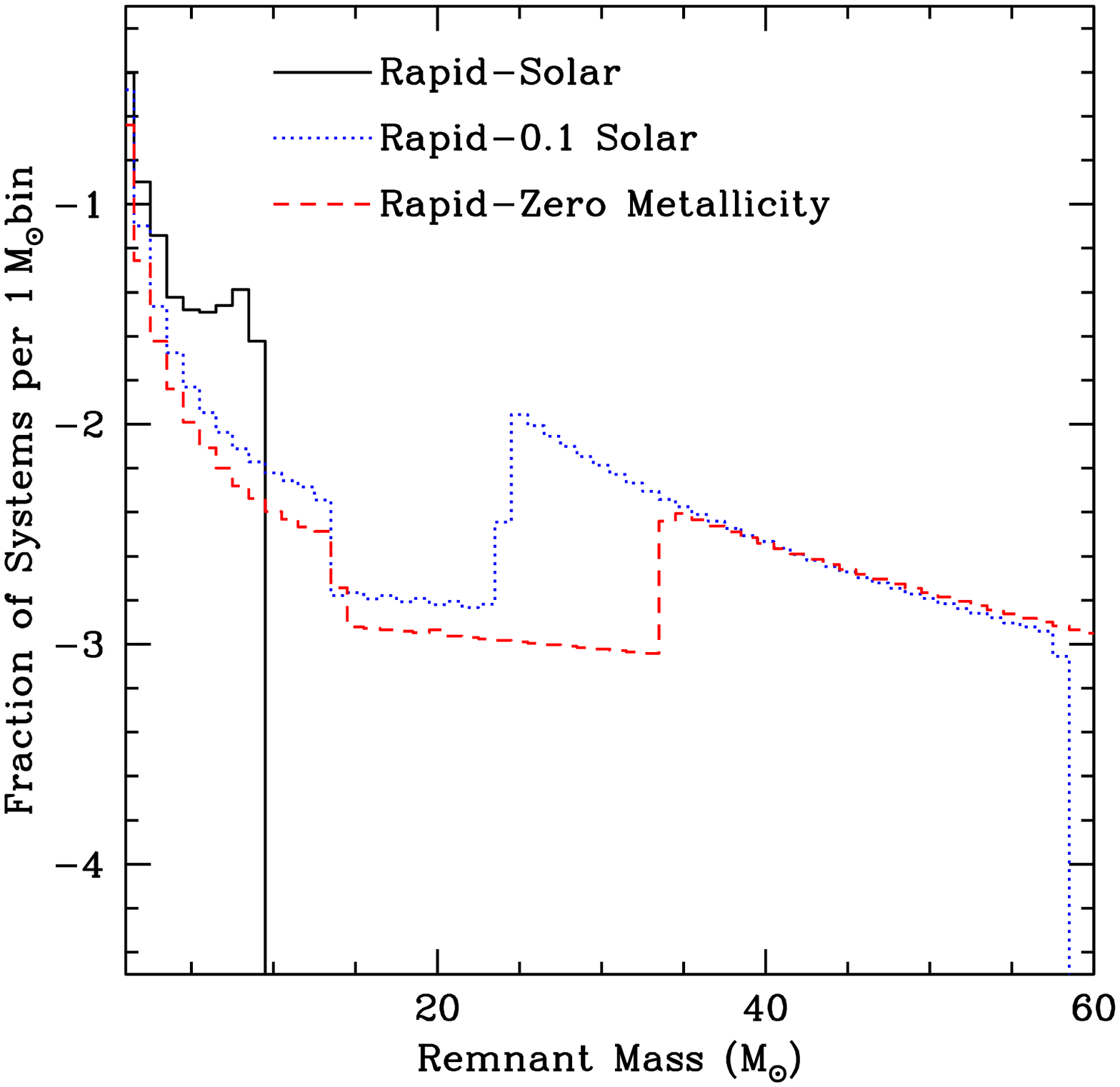}
\caption{Distribution of NS and BH remnant masses from our delayed
  explosion model for a range of metallicities.  Metallicity
  determines the maxmimum BH mass with a sharp peak at the
  maximum mass at collapse.}
\label{fig:distmetal}
\end{figure}

With our initial/remnant mass relation, and assuming a value for the
initial mass function (IMF), we can now derive mass distributions for
both NSs and BHs as a function of metallicity.
These mass distributions for a Salpeter ($\alpha=2.35$) IMF are
shown in figure~\ref{fig:distmetal}.  At lower metallicities, the
number of low mass NSs increases. This is because the lower
limit for stellar collapse and NS formation decreases with
decreasing metallicity.  The maximum BH mass increases
with decreasing metallicity, as mass-loss becomes weaker, leading to
more massive stars, and hence more massive remnants.

\begin{figure}
\plottwo{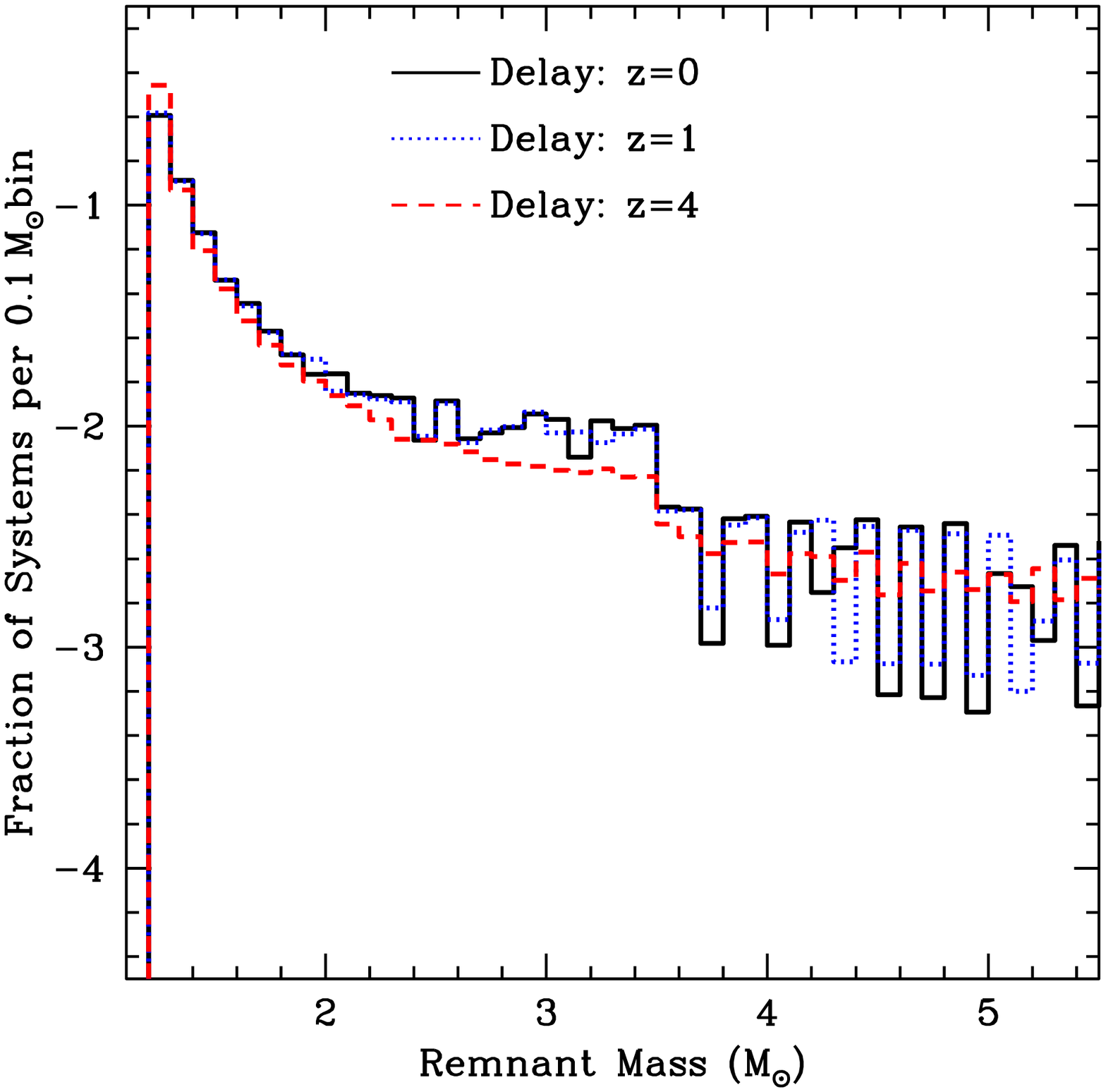}{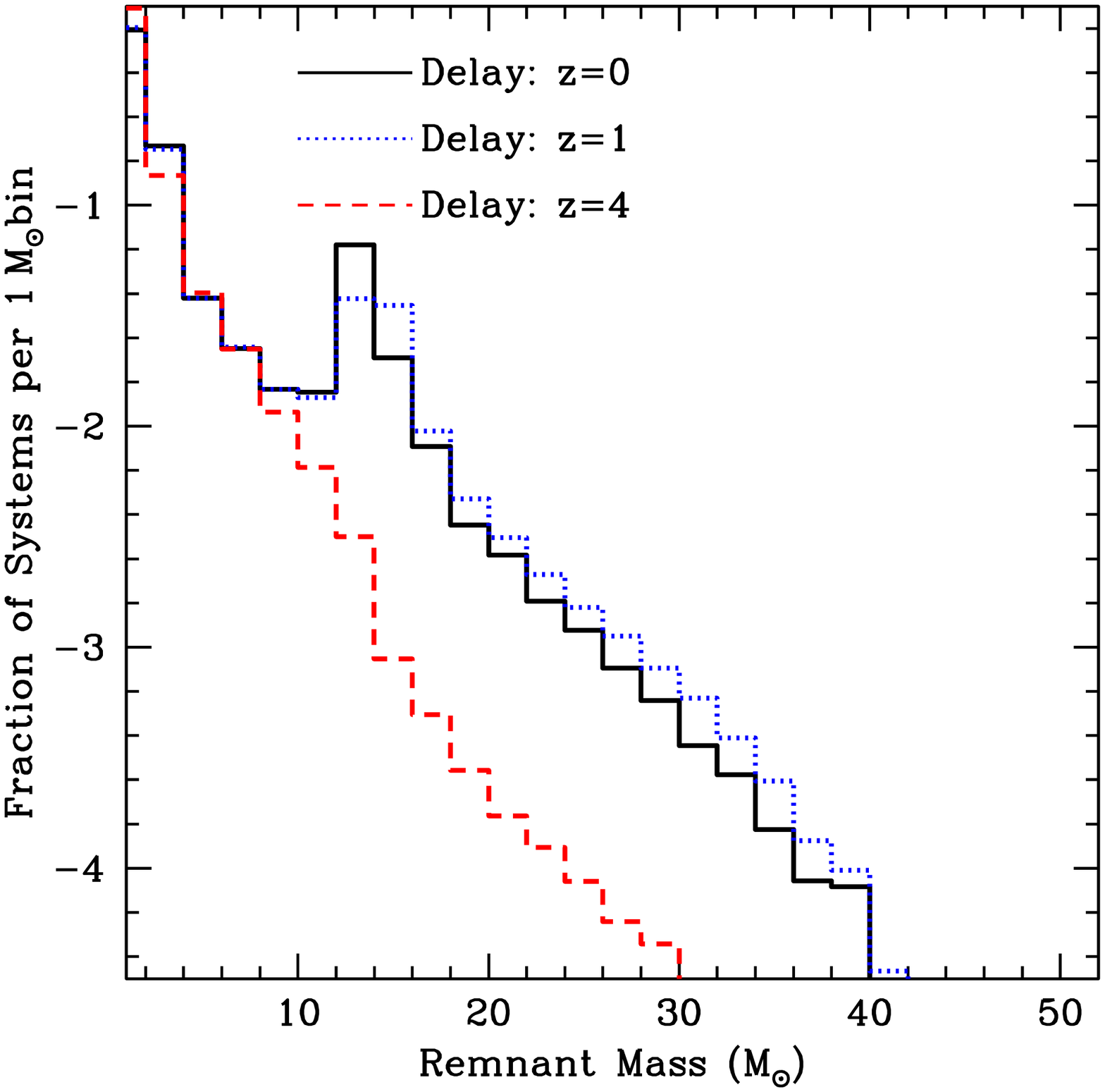}
\caption{Distribution of NS and BH remnant masses from our delayed
  explosion model as a function of redshift.  The lower metallicities
  at higher redshifts lead to more mass BH remnants.}
\label{fig:distred}
\end{figure}

\begin{figure}
\plottwo{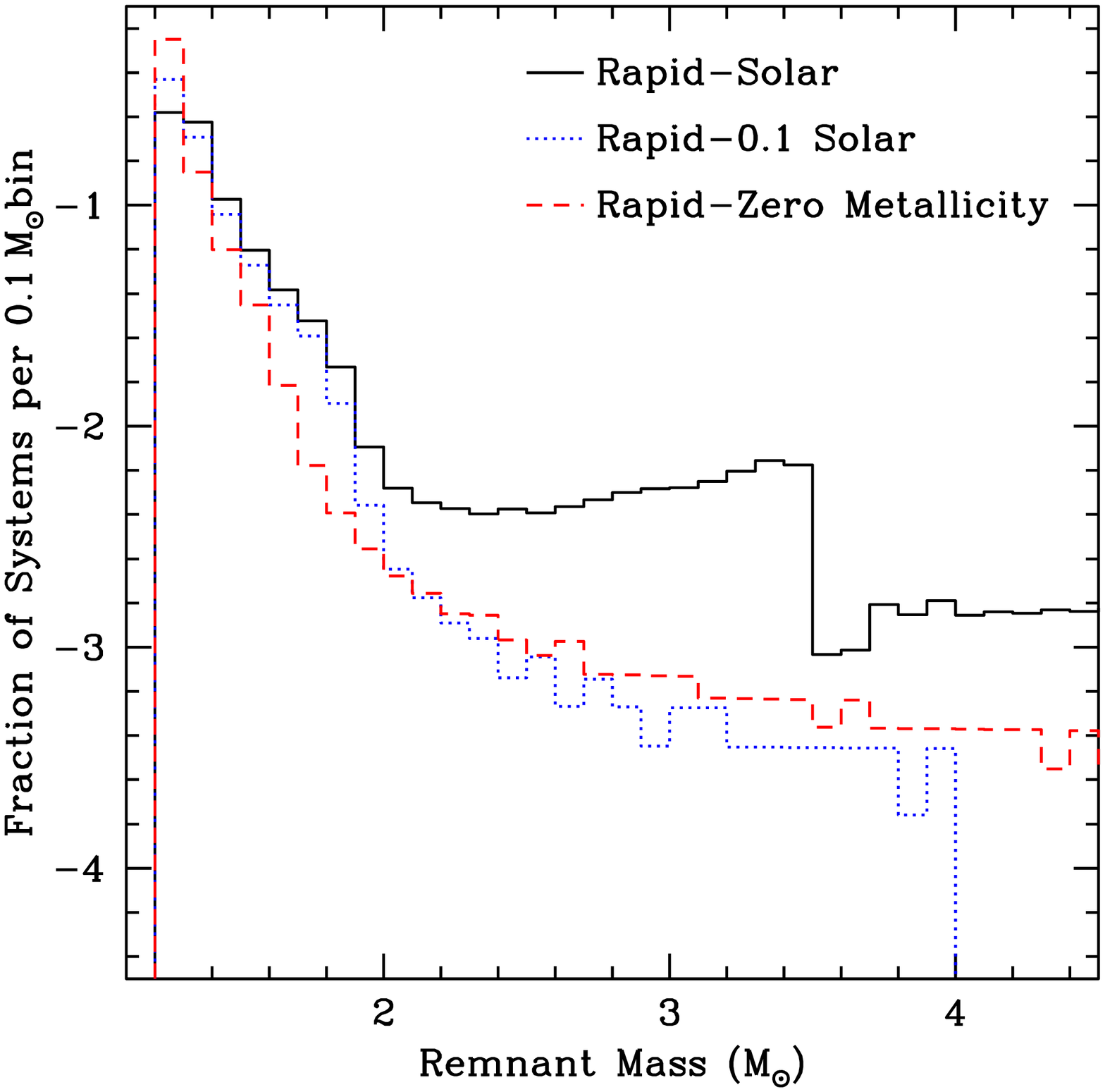}{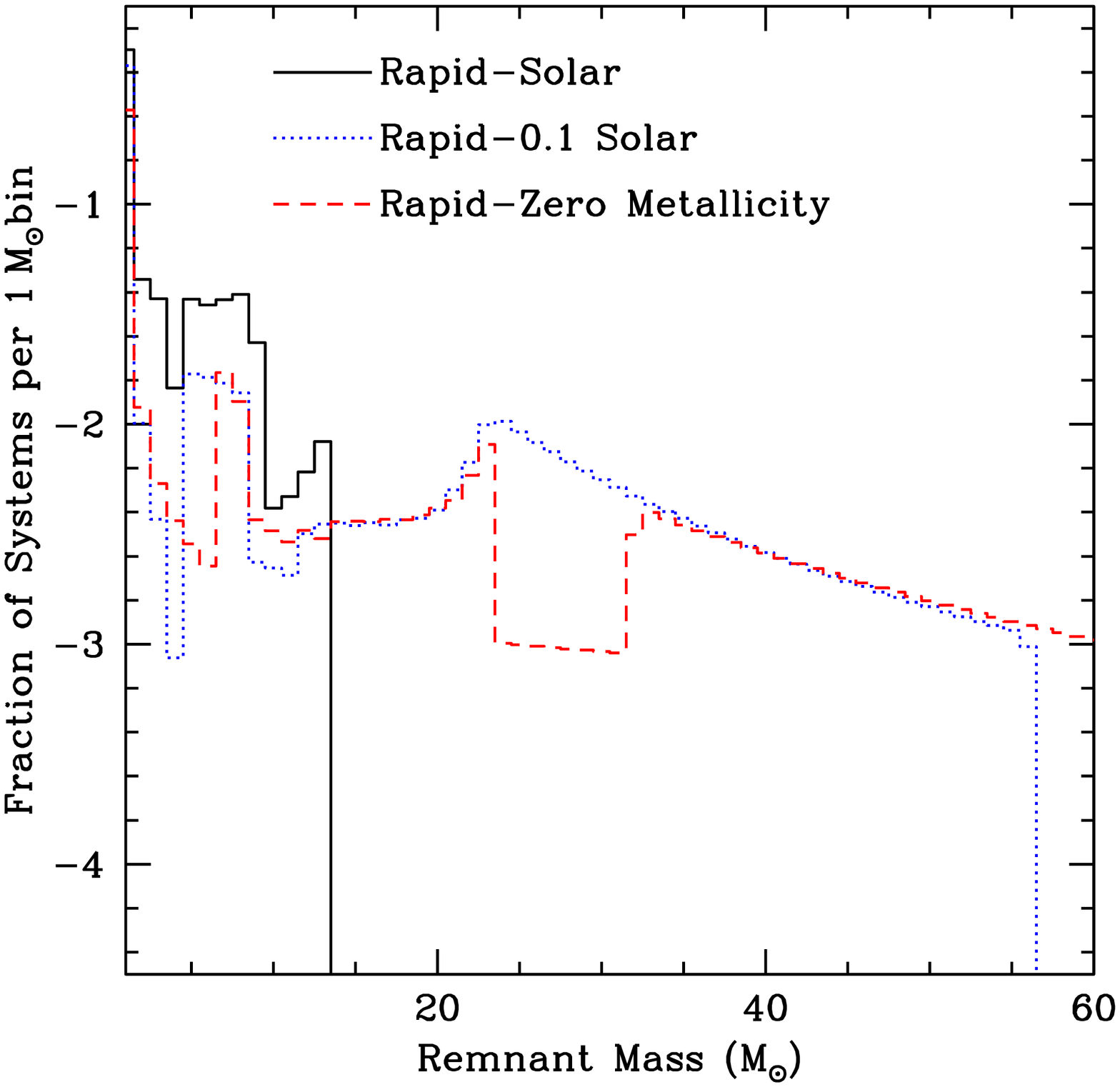}
\caption{Distribution of NS and BH remnant masses from our rapid
  explosion model for a range of metallicities.  Metallicity
  determines the maxmimum BH mass with a sharp peak at the
  maximum mass at collapse.}
\label{fig:distmetalrapid}
\end{figure}

\begin{figure}
\plottwo{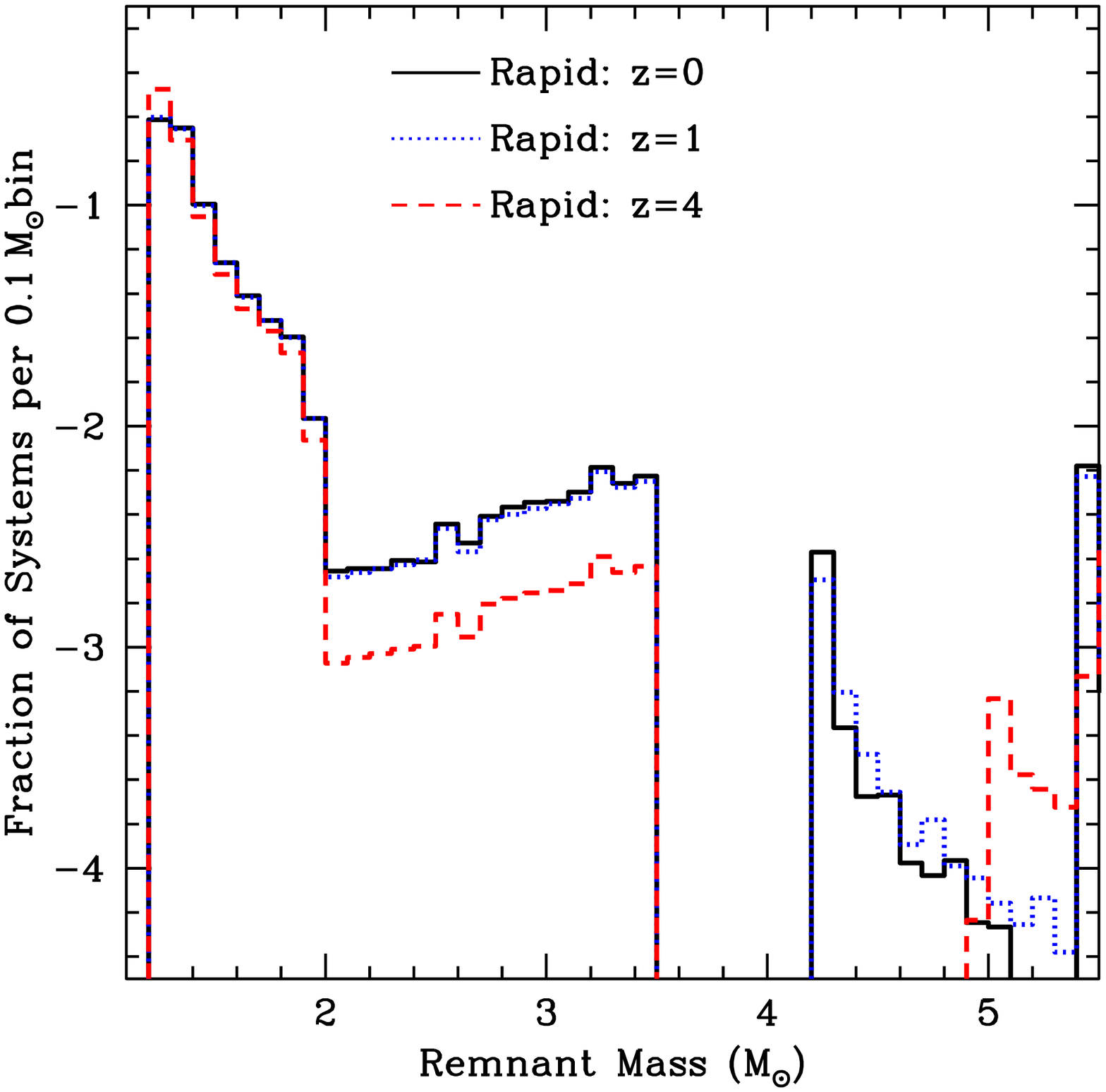}{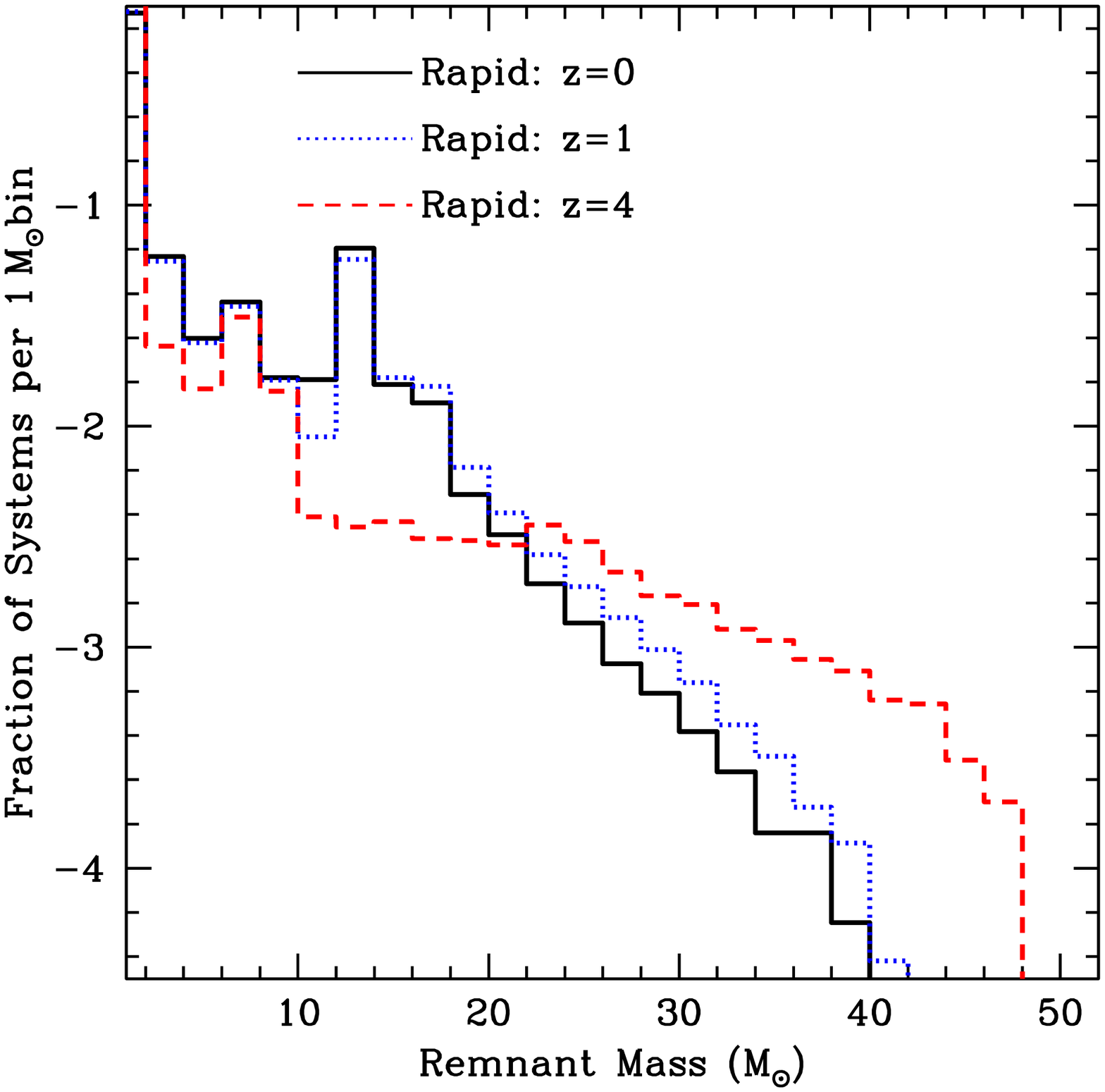}
\caption{Distribution of NS and BH remnant masses from our rapid
  explosion model as a function of redshift.  The lower metallicities
  at higher redshifts lead to more mass BH remnants.}
\label{fig:distredrapid}
\end{figure}

Using the metallicity distribution as a function of redshift (Young \&
Fryer 2008), we can determine mass distributions of NSs and BHs as a
function of redshift (Fig. ~\ref{fig:distred}).  The corresponding
distributions versus redshift and versus metallicity for our rapid
explosion models are shown in figures~\ref{fig:distmetalrapid}
and~\ref{fig:distredrapid}.  The rapid explosion produces very few
compact remnants in the gap region: for the solar metallicity models
of Woosley et al. (2002), $\lesssim$18\% of all remnants above
2\,M$_\odot$ lie in the 2--5\,M$_\odot$ gap region.  At 0.1 solar
metallicity, this fraction decreases to less than 1.5\%.  The Limongi
\& Chieffi (2006) models, with their lower mass-loss, predict this
small fraction of remnants at solar metallicity.  The BH mass
distribution peaks at roughly 7--8\,M$_\odot$.  On the other hand, the
delayed explosion model produces far more remnants in the gap region:
33\% at solar metallicity, and 17\% at 0.1 solar metallicity.  If the
observed gap represents the true compact object formation mass
distribution, the rapid explosion engine is a better match to the
observed gap in remnant masses between
2--5\,$M_\odot$~\citep{bailyn98, ozel10}.  To directly compare with
these observations, we must focus on X-ray binary systems (see Belcyzinski et al. 2011).

\subsection{Sources of Uncertainty}
\label{sec:caveats}

For remnant masses below 11\,M$_\odot$, we have two sources of error.
First, uncertainties in modeling mass-loss and convection (dredge-up)
can alter the lower limit for NS formation.  We have assumed that at
zero metallicity this lower limit drops to 6.3\,M$_\odot$, but a limit
of 8\,M$_\odot$ also fits within the errors of the calculations
(Poelerands et al. 2008).  Second, differences in the results of
core-collapse models lead to an error in the mass of the NSs, allowing
a range from 1.28--1.36\,M$_\odot$.  Observations placing constraints 
on the lowest neutron star mass could reduce this error.

For remnant masses produced by stars above 11\,M$_\odot$, the
uncertainties are significantly larger. Our models are based on the
assumption that the standard convective engine is the correct engine
for most supernovae.  We assume the maximum efficiency for ejecting
stellar material (i.e. the entire explosion energy is used to unbind the outer
layers of the star).  This assumption naturally leads to an underestimate
of the final remnant mass.  However, we also assume that the proto-neutron
star lacks an energy reservoir, despite the fact that there are many mechanisms by which it
could inject energy and affect the fallback.  For example, proto-NS winds could generate
additional energy in the explosion, thereby reducing fallback and decreasing
the remnant mass.  Even more
important, Fryer et al. (1996, 2006) and Fryer (2008) found that
fallback material will drive outflows, self-limiting the total amount of
fallback. As we discussed, other magnetic fields can also drive 
outflows, preventing fallback (e.g. Maeda et al. 2007, Kasen \&
Bildsten 2010).  All of these effects would drive our final remnant
mass lower, and ignoring them overestimates the remnant mass.

\begin{figure}
\plotone{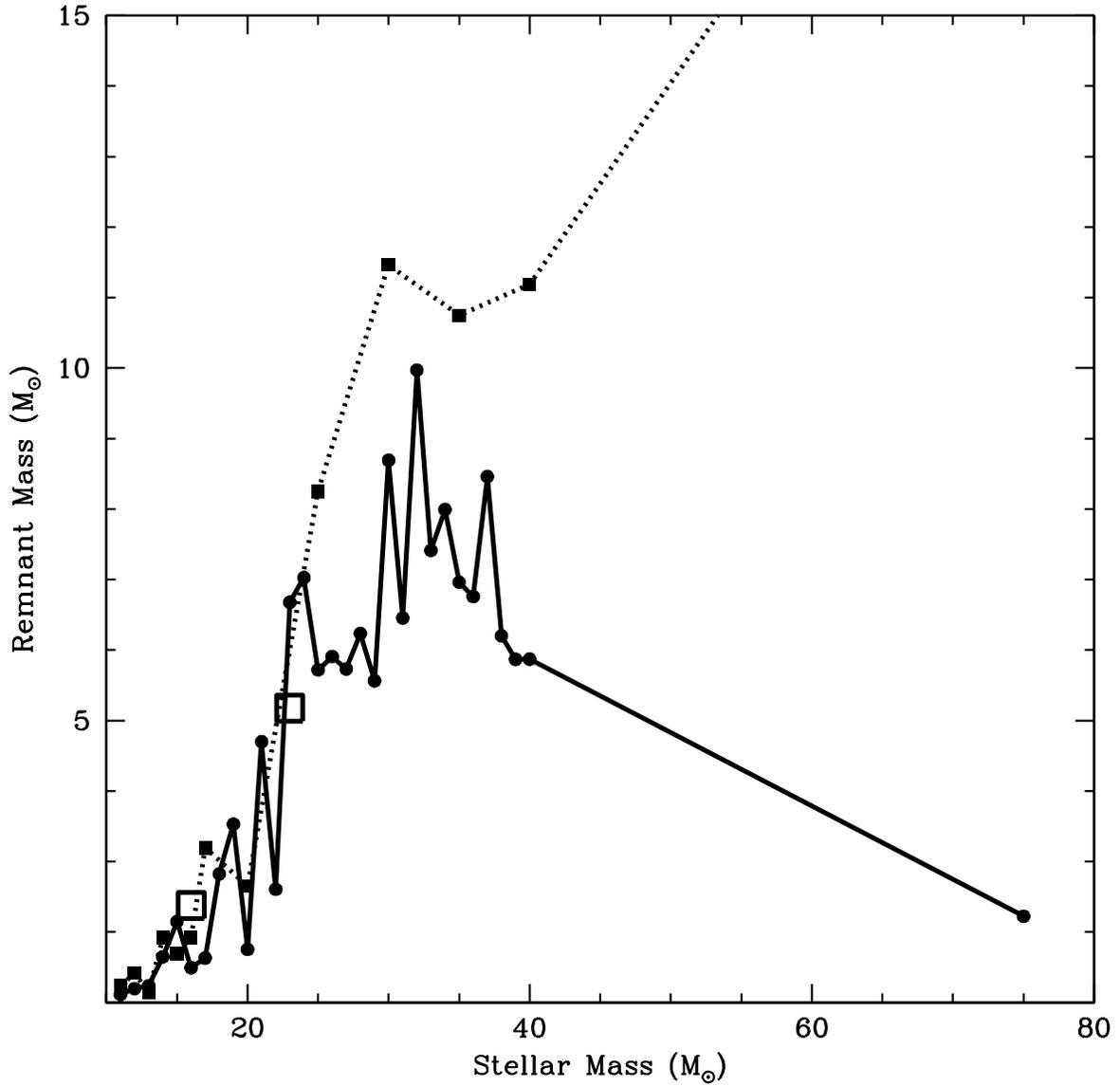}
\caption{Remnant mass versus ZAMS mass for 3 different suites of
  calculations: Woosley et al. (2002) solar metallicity (solid line),
  Limongi and Chieffi (2006) solar metallicity (dotted), and the
  binary models of Young et al. (2008) at solar metallicity (open
  squares).  Note that the effects of binaries below 25\,M$_\odot$
  does not change the remnant mass whatsoever.}
\label{fig:stellarcomp}
\end{figure}

In addition, the mass distribution is sensitive both to the
prescription used for mass loss in the stellar evolution codes and the
stability of the stellar evolution codes.
Figure~\ref{fig:stellarcomp} compares the results from three
different sets of stellar models using three different stellar evolution
codes: Woosley et al. (2002) solar metallicity (solid line), Limongi
and Chieffi (2006) solar metallicity (dotted), and the binary models
of Young et al. (2008) at solar metallicity (open squares).  

Different stellar models utilize different prescriptions for mass
loss, accounting for the huge differences between
the Limongi \& Chieffi (2006) and Woosley et al. (2002) results
above 30\,$M_\odot$.  Much of this difference is just a scaling
factor: the primary difference in the mass-loss prescription
is the coefficient in front of the metallicity-dependent term.
For example, the Limongi \& Chieffi (2006) solar metallicity models
are very similar to the Woosley et al. (2002) models at $1/3$ solar
metallicity.  To determine the remnant mass distribution of the
Limongi \& Cheiffi (2006) models, we would use our remnant mass
prescriptions
(e.g. equations~\ref{eq:massive},~\ref{eq:massiverapid}) 
at $1/3$ solar metallicity.

The jagged nature of our calculated mass estimate comes from
instabilities in the mixing within the stellar evolution code.
Because of this, the core masses are not a smooth function of
progenitor mass.  The high remnant mass at $23\,M_\odot$ in our rapid
explosion models in Fig.~\ref{fig:mremwooR} is a demonstration of how
important this mixing can be. It is possible that better convection
models will reduce the non-smooth dependence of core mass on
progenitor mass (see Young et al. 2005), but this remains to be
established.

We have also neglected the effects of binaries in the discussion above.  Because fallback
occurs prior to the shock hitting the hydrogen envelope, and because
the hydrogen envelope has very little binding energy, the primary
effect of binaries on the mass distribution will be to reduce
the maximum mass of the remnant to that of the helium core mass. 
Fryer \& Kalogera (2001) have studied this effect, and find that it is
negligible in the case of NSs.  However, the broader remnant mass distribution
is sensitive to binary effects, and we  
study these next (section~\ref{sec:binary}).

\section{Prescriptions for Compact Object Formation in Binaries}
\label{sec:binary}

In this section we describe three approaches to computing the mass of a compact
object based on different core collapse/supernova simulations. Each
method can be easily employed in analytic or population synthesis approaches,
provided that the mass of a star and its CO core mass
are known at the time of supernova explosion.  The prescriptions
provide a mass of a remnant compact object without distinguishing
between the type (NS or BH).  However, in our standard approach
we adopt a maximum NS mass of $M_{\rm NS,max}=2.5 \msun$;
the lack of observations of compact remnants with masses between 
$\sim 2$--$3.5 \msun$ and theoretical uncertainties make it 
difficult to determine an exact neutron star mass.
To present typical initial-remnant mass relations we use the Hurley et al.
(2000) formulae as implemented in the {\tt StarTrack} population
synthesis code (Belczynski et al. 2002, 2008), with updated wind mass
loss rates (Belczynski et al.  2010), to obtain the properties of a
star at the time of supernova explosion.  In addition, we
allow for the formation of NSs via electron-capture
supernovae (ECS) (see below).

\subsection{StarTrack}

Here we present a brief description of the computation of compact object masses 
in the {\tt StarTrack} population synthesis code (for the full description see
Belczynski et al. 2008, 2010). 

We determine the mass of a NS/BH remnant using information on the final CO
mass $M_{\rm CO}$ combined with the knowledge of the pre-supernova mass of the 
star $M$. For a given initial ZAMS mass, the final CO core mass is obtained from 
the original Hurley et al. (2000) formula, while we use the models of Timmes, 
Woosley, \& Weaver (1996)m with the addition of Si shell mass, to 
estimate final FeNi core mass (which we here refer to as proto-compact object 
mass). The proto compact object mass is obtained from:\\ 
\begin{equation}
 \left\{ \begin{array}{rll}
M_{\rm proto}= &  1.50 \msun & M_{\rm CO} < 4.82 \msun\\
M_{\rm proto}= &  2.11 \msun & 4.82 \leq M_{\rm CO} < 6.31 \msun\\
M_{\rm proto}= &  0.69 M_{\rm CO} - 2.26 \msun & 6.31 \leq M_{\rm CO} < 6.75 \msun\\ 
M_{\rm proto}= &  0.37 M_{\rm CO} - 0.07 \msun & M_{\rm CO} \geq 6.75 \msun.
\end{array}
\right.
\label{eq1}
\end{equation}
The fallback of material after the launch of the explosions adds mass
to the remnant.  To calculate the amount of fallback, $M_{\rm fb}$,
for a given core mass, $M_{\rm CO}$, we employ:
\begin{equation}
  \left\{ \begin{array}{rll}
     M_{\rm fb} = & 0 \msun & M_{\rm CO} < 5.0 \msun\\
     f_{\rm fb} = & 0.378 M_{\rm CO}-1.889 \msun & 5.0 \leq M_{\rm CO} < 7.6 \msun \\
     f_{\rm fb} = & 1.0 & M_{\rm CO} \geq 7.6 \msun 
\end{array}
\right.
\end{equation}
with $M_{\rm fb}=f_{\rm fb} (M-M_{\rm proto})$ in the mass range for which the
fractional fall back $f_{\rm fb}$ is given\footnote{The formula for the 
fall back presented here is somewhat different than in Belczynski et al.
(2008), however it results in very similar fallback values. The ranges for 
partial fall back ($0<f_{\rm fb}<1$) and direct BH formation ($f_{\rm fb}=1$) 
are estimated from core collapse models of Fryer, Woosley, \& Hartmann (1999) 
and the analysis of Fryer \& Kalogera (2001).}

The final remnant (baryonic) mass is calculated from 
\begin{equation}
M_{\rm rem,bar} = M_{\rm proto} + M_{\rm fb},
\end{equation}
and we convert baryonic to gravitational mass ($M_{\rm rem}$) using:
\begin{equation}
M_{\rm rem,bar} - M_{\rm rem} = 0.075\ M_{\rm rem}^2
\label{eq3}
\end{equation}
for NSs (Lattimer \& Yahil 1989; see also Timmes et al. 1996), while 
for BHs we simply approximate the gravitational mass with 
\begin{equation}
 M_{\rm rem} = 0.9\ M_{\rm rem,bar}.
\label{eq4}
\end{equation}

\subsection{Rapid Supernova Mechanism}

To calculate the final mass of a compact object we need to know the mass of a
star, $M$, and its CO core mass, $M_{\rm CO}$, at the time of core collapse/SN explosion.  
For the rapid explosion mechanism, the explosion either occurs quickly or not at all.  
We set the proto-compact object mass
\begin{equation}
M_{\rm proto} = 1.0 \msun,
\end{equation}
independent of exploding star mass according to hydrodynamical
simulations of SN explosions~\citep{Woo02}.  Depending on the amount of
mass above the proto-compact object ($M-M_{\rm proto}$) and the strength of
the explosion, potential fall back may increase the mass of the compact object.
We calculate the amount of fall back, $M_{\rm fb}$, for a given core mass,
$M_{\rm CO}$:\\ 
\begin{equation}
  \left\{ \begin{array}{rll}
     M_{\rm fb} = & 0.2 \msun & M_{\rm CO} < 2.5 \msun\\
     M_{\rm fb} = & 0.286 M_{\rm CO}-0.514 \msun & 2.5 \leq M_{\rm CO} < 6.0 \msun \\
     f_{\rm fb} = & 1.0  & 6.0 \leq M_{\rm CO} < 7.0 \msun \\
     f_{\rm fb} = & a_1 M_{\rm CO} + b_1  & 7.0 \leq M_{\rm CO} < 11.0 \msun\\
     f_{\rm fb} = & 1.0 & M_{\rm CO} \geq 11.0 \msun 
\end{array}
\right.
\end{equation}
with $a_1=0.25 - {1.275 \over M-M_{\rm proto}}$, $b_1=-11 a_1+1$ and
$M_{\rm fb}=f_{\rm fb} (M-M_{\rm proto})$ in the mass range for which $f_{\rm fb}$
is given. 

The final baryonic mass of the remnant is then
\begin{equation}
M_{\rm rem,bar} = M_{\rm proto} + M_{\rm fb}
\end{equation}
and the gravitational mass of the remnant is obtained from eqs.~\ref{eq3} and~\ref{eq4}.

\subsection{Delayed Supernova Mechanism}
The calculation of the final mass of a compact object for the delayed mechanism
is similar to the rapid SN case detailed above. The formulae are based on 
the delayed SN (SASI) calculations discussed in earlier sections. 
First we calculate the proto-compact object mass
\begin{equation}
  \left\{ \begin{array}{rll}
    M_{\rm proto} =&  1.2 \msun & M_{\rm CO} < 3.5 \msun\\
    M_{\rm proto} =&  1.3 \msun & 3.5 \leq M_{\rm CO} < 6.0 \msun \\
    M_{\rm proto} =&  1.4 \msun & 6.0 \leq M_{\rm CO} < 11.0 \msun \\
    M_{\rm proto} =&  1.6 \msun & M_{\rm CO} \geq 11.0 \msun
\end{array}
\right.
\end{equation}
This assumes that the delay increases for more 
massive cores, causing more material to accrete onto the proto-neutron 
star during the explosion.

The amount of fall back is given by:\\ 
\begin{equation}
  \left\{ \begin{array}{rll}
     M_{\rm fb} = & 0.2 \msun & M_{\rm CO} < 2.5 \msun\\
     M_{\rm fb} = & 0.5 M_{\rm CO}-1.05 \msun & 2.5 \leq M_{\rm CO} < 3.5 \msun \\
     f_{\rm fb} = & a_2 M_{\rm CO} + b_2  & 3.5 \leq M_{\rm CO} < 11.0 \msun\\
     f_{\rm fb} = & 1.0 & M_{\rm CO} \geq 11.0 \msun 
\end{array}
\right.
\end{equation}
with $a_2=0.133 - {0.093 \over M-M_{\rm proto}}$, $b_2=-11 a_2 + 1$ and
$M_{\rm fb}=f_{\rm fb} (M-M_{\rm proto})$ in the mass range for which $f_{\rm fb}$
is given. 

The final baryonic mass of the remnant is then
\begin{equation}
M_{\rm rem,bar} = M_{\rm proto} + M_{\rm fb},
\end{equation}
and the gravitational mass of the remnant is obtained from eqs.~\ref{eq3} and
~\ref{eq4}.

\subsection{Electron Capture Supernovae}

In our calculations for the masses of compact objects, as detailed above, 
we allow for NS formation through electron-capture supernovae
(ECS, e.g., Podsiadlowski et al. 2004), following the approach in Belczynski et
al. (2008).  From Hurley et 
al. (2000) we use the He core mass at the AGB base to set the limit
for the formation of various CO cores. If the He core mass is smaller
than $M_{\rm cbur1}$ the star forms a degenerate CO core, and ends up
forming a CO WD. If the core is more massive than $M_{\rm cbur2}=2.25
\msun$ the star forms a non-degenerate CO core with subsequent burning
of elements until the formation of a FeNi core which ultimately
collapses to a NS or BH.  Stars with cores between $M_{\rm cbur1}$
and $M_{\rm cbur2}$ may form partially degenerate CO cores. If such a
core reaches a critical mass ($M_{co,crit}=1.08 \msun$, Hurley et
al. 2000), it ignites CO off-center and non-explosively burns CO into
ONe, forming a degenerate ONe core.  If in subsequent evolution the
ONe core increases its mass to $M_{\rm ecs}=1.38 \msun$ the core
collapses due to electron capture on Mg, and forms a NS. We refer
to these as ECS NS, to distinguish them from NS from regular iron
core collapse. The ECS NSs are assumed to
have unique masses of $M_{\rm rem,bar}=M_{\rm ecs}$. If the ONe core
mass remains below $M_{\rm ecs}$ the star forms a ONe WD.

Hurley et al. (2000) suggested $M_{\rm cbur1}=1.66 \msun$
corresponding to $M_{\rm zams}=6.5 \msun$ for $Z=0.02$. Later
calculations with an updated evolutionary code (Eldridge \&
Tout 2004a,b) indicated that ECS may occur for higher initial masses
($M_{\rm zams} \gtrsim 7.5 \msun$). For our standard model we adopt
$M_{\rm cbur1}=1.83 \msun$ ($M_{\rm zams}=7.0 \msun$), and this
results in ECS NS formation above $M_{\rm zams}=7.6 \msun$ (for masses
$M_{\rm zams}=7.0-7.6 \msun$ the ONe core does not reach $M_{\rm ecs}$
and a ONe WD is formed).  It is noted that binary evolution through
Roche-Lobe Overflow may either decrease the initial mass of the ZAMS star
required to form the core of mass $M_{\rm cbur1}$ (due to
rejuvenation) or increase it (due to mass loss). Therefore, binary
evolution effectively leads to wider initial progenitor masses for ECS
NS formation. Metallicity and wind mass loss may also influence the ECS
NS formation range.

\subsection{Natal Kicks}

At the time of birth NSs and BHs receive a natal kick due to
asymmetries in the SN explosions. We use the distribution inferred from observed 
velocities of radio pulsars by Hobbs et al. (2005): a single Maxwellian with 
$\sigma=265$ km sec$^{-1}$. 

Compact objects formed without fall back receive full kicks drawn from
the above distribution. This is the case for most NSs, with the
exception  of ECS  NSs, for which we adopt either no 
natal kicks (standard model) or full kicks (to explore a range of parameters). 
The same approach is applied to NSs formed via accretion induced collapse of WD to NS 
in an accreting binary system. 

Additionally, compact objects formed with small amounts of fall back 
($M_{\rm fb} < M_{\rm fb,small}$) receive full kicks (this includes
some low mass BHs). In our standard approach we adopt 
$M_{\rm fb,small} = 0.0 \msun$ for the {\tt StarTrack} scheme, $M_{\rm
fb,small} = 0.2 \msun$ for the rapid and delayed SNe. 
However, to fully explore parameter space we also consider full kicks with a broad
fall back mass range (as high as $M_{\rm fb,small} \sim 1 \msun$).

For compact objects formed with noticeable fall back ($M_{\rm fb} \geq 
M_{\rm fb,small}$), kicks are lowered proportional to the amount of fall back:
\begin{equation}
V_{\rm kick} = (1-f_{\rm fb}) \sqrt{V_{\rm x}^2 + V_{\rm y}^2 + V_{\rm z}^2},
\label{kick001}
\end{equation}
where $V_{\rm x}, V_{\rm y}, V_{\rm z}$ are the three velocity components drawn from 
Hobbs et al. (2005) distribution, and $f_{\rm fb}$ is 
the fraction (from 0 to 1) of the stellar envelope that falls back.

For the most massive BHs, which are formed silently (without a SN explosion) via direct
collapse ($f_{\rm fb}=1$) of a massive star, we assume that no
natal kicks are imparted.  In this formalism, the kick is solely a function 
of the fall back fraction ($f_{\rm fb}$), and not of the mass of the compact 
remnant. 

\subsection{Prescription Results}
\label{relations}

The mass distributions from our prescriptions for remnant
masses at solar metallicity are shown in Figure~\ref{Mrem2}.  The
dashed line shows the results from a standard StarTrack calculation
(e.g. Belczynski et al. 2010), the solid line shows the
results from our delayed engine, and the dot-dashed line shows the
results for the rapid engine.  The rapid explosion 
will produce many more BHs of $\sim 8\,M_\odot$ than the 
delayed explosion. For intermediate mass stars that produce BHs 
($M_{\rm zams} \approx 40$--$80 \msun$) the StarTrack BHs are more massive 
than BHs predicted in more recent simulations (for both delayed and rapid models). 

Most interesting are the differences between these models in 
producing compact remnants below $5\,M_\odot$.  The rapid 
explosion mechanism produces a range of compact remnant masses 
up to $2\,M_\odot$, but makes very few
remnants of mass in the range $\sim 2$--$5\,M_\odot$.  The StarTrack models
partially fill this gap, while the delayed explosion models conribute a
significant population of remnants in the gap. These 
trends persist at lower metallicities (Fig.~\ref{Mrem1}). 

This striking gap in compact object mass ($\sim 2$--$5\,M_\odot$) results
from the sensitivity of the rapid models to the structure of the core.  
In the Woosley et al. (2002) and Woosley \& Heger (2007) models, enhanced 
burning in the late evolutionary stages leads to sharply different internal 
structures of stars with mass $M_{\rm zams} \approx 20$--$25 \msun$. The
internal structure in turn affects the supernova explosion and dramatically
changes the mass of the remnant formed in the collapse.
The changes are such that in this mass range we get 
an abrupt transition of compact object mass from $2 \msun$ 
to much higher mass, with no compact objects in between (see Figs.~\ref{Mrem2} 
and ~\ref{Mrem1}) .  
This result agrees with studies of O'Connor \& Ott (2011), who find that
the bounce compactness changes dramatically for these stars, altering
their fate. The ultimate fate of these stars is sensitive to the
late-stage burning, and if the burning changes with newer stellar 
evolution codes, the remnant mass distribution will change commensurately.

\section{Conclusions and Observational Comparisons}

We find that different supernova engines produce very different remnant mass
distributions. We now review the observational constraints, and make comparisons
with our predictions.

\subsection{X-ray Binary Remnant Mass Distribution}

One of the strongest observational constraints is the gap in BH
mass between 2--5\,M$_\odot$~\cite{Farr11,ozel10}.  However, these
studies do not include systems which do fall in the gap, such as
4U1700-37\citep{Clark02} with a measured mass of
$2.44\pm0.27$\,M$_\odot$.  We expect some remnants to fall in the gap, but have
difficulty estimating the number. Our model predictions range from
$\sim$15\% for our rapid collapse to $\sim$40\% for our delayed model in
the 2--4\,M$_\odot$ mass range (section 4.4).  The high space velocity
of 4U1700-37 points out another difficulty in using X-ray binary mass
observations to constrain the remnant mass of supernovae: BH
and NS kicks.

Many of the proposed NS kick mechanisms will also impart
kicks onto BHs at formation.  To impart a kick on the compact
remnant from stellar collapse, momentum conservation requires asymmetries.
Most compact remnant kick mechanisms either argue for asymmetries in the
baryonic ejecta from the supernova or in the neutrino emission.  Here
we review the implications for compact remnant kicks for our two
explosion models utilizing these two kick mechanisms.

For ejecta driven-kicks, no kick is imparted on BHs formed without
supernova explosions (since there is then no ejecta).  Low-mass BHs,
on the other hand, are formed when the supernova explosion is delayed
and the explosion is weak.  If the convection in the supernova engine
develops low-mode instabilities, the resulting explosion can be
asymmetric~\citep{Her92, Buras03, Blondin03}.  The SASI engine is
characterized by low-mode convection and several groups have argued
that these large asymmetries will drive strong
kicks~\citep{Buras03,Blondin03,Scheck06,Wongwathanarat10}; see Fryer
\& Young (2007) for a different interpretation of the low-mode
convenction.  If the low mode convection in the SASI engine produces
strong kicks, the explosions with the longest delays will have the
largest ejecta asymmetries and, hence, the strongest kick momenta.
But these delayed supernova engines also produce the weakest supernova
explosions, forming massive NSs and low-mass BHs.  We thus expect
massive NSs and low-mass BHs to have the highest momentum ``kick'' of 
any compact remnant.

For the rapid explosion model most BHs are formed without 
a supernova explosion, and only the small fraction of low-mass black 
holes will receive kicks.  This is to be contrasted with the delayed model,
where a much larger 
fraction of BHs are formed from fall back, and these low-mass 
BHs will have large kicks.

Likewise, for most asymmetric neutrino kick mechanisms, the longer the
delay in the supernova engine, the stronger the momentum
asymmetry~\citep{Socrates05, Kusenko04, FryerKusenko06}.  Again,
massive NSs and low mass BHs receive the 
largest kicks of any of our compact remnants under our delayed explosion
model. 

Strong kicks unbind binary systems.  If strong kicks occur in
high-mass NSs and low-mass BHs, these systems 
are more likely to be disrupted in the massive star collapse.  This 
bias will make it difficult to use remnant mass distributions 
derived from X-ray binary systems without detailed population 
synthesis models.

\subsection{Implications for Supernovae}

Although our focus has been on estimating the compact remnant mass
distribution, our models also make predictions for the distribution of
explosion energies for core-collapse supernovae.  The explosion energy
distributions for our two convection-enhanced engines at both solar
and zero metallicity are shown in figure~\ref{fig:distexpe}.  We have
assumed that core-collapse supernovae from low-mass stars and electron
capture supernovae all produce the same low-energy,
$0.55\times10^{51}$\,erg explosion. Since the minimum mass decreases
with metallicity, these supernovae dominate the explosions at low-metallicity,
even with the relatively shallow Salpeter IMF ($\alpha=2.35$).  

From figure~\ref{fig:distexpe} we note that, except for the low mass
supernovae such as electron capture supernovae, our rapid explosion
model produces only strong explosions (above $10^{51}$\,erg). The
delayed model produces a broader range of explosion energies, with peak
energies below that of the rapid explosion model.  Even at solar
metallicity less than 50\% of the explosions from the delayed
engine have energies in excess of $10^{51}$\,erg.  In principal the
distribution of supernova explosion energies, if sufficiently
accurate, could distinguish between our two engines.  For example, a
low-energy explosion from a massive star is almost certainly arising
from a delayed explosion.  Similarly, any explosion above
$1.5\times10^{51}$\,erg produced by a slowly rotating star is likely
to result from a rapid explosion.

It is likely that a combination of both the delayed and rapid explosion engines
will be required to explain supernovae, and future observations may determine the relative
fraction.

\subsection{Implications for Gravitational Waves}

Gravitational waves are expected to provide new constraints on the
formation of compact remnants.  We use
the {\tt StarTrack} population synthesis code, incorporating the binary
prescriptions discussed above,  to calculate delayed (e.g. SASI) and rapid models
for the initial--remnant mass relation.
Additionally, we employ two models of common envelope evolution.
In the first (model A) we employ the standard energy balance
(Webbink 1984) to calculate the outcome of common envelope evolution with 
a Hertzsprung gap donor.
In the second model (B) we
assume that, when a Hertzsprung gap star fills its Roche lobe and
starts the common envelope phase, it merges with its companion and ends
its binary evolution.  The same energy balance is used to compute the common
envelope for
any type of donor in both models. Since many BH-BH progenitors
experience common envelope evolution with a Hertzsprung gap donor, the birth and merger
rates of BH-BH binaries are greatly reduced in model B (for details and
the physical motivation see Belczynski et al. 2007). We consider two 
metallicities $Z_{\odot}=0.02$ and $0.1 Z_{\odot}=0.002$ (approximately
covering the range of metallicity observed in local Universe).
We focus on binary stars that form double BH systems (BH-BH) at the 
end of their evolution. 

\begin{figure}
\plotone{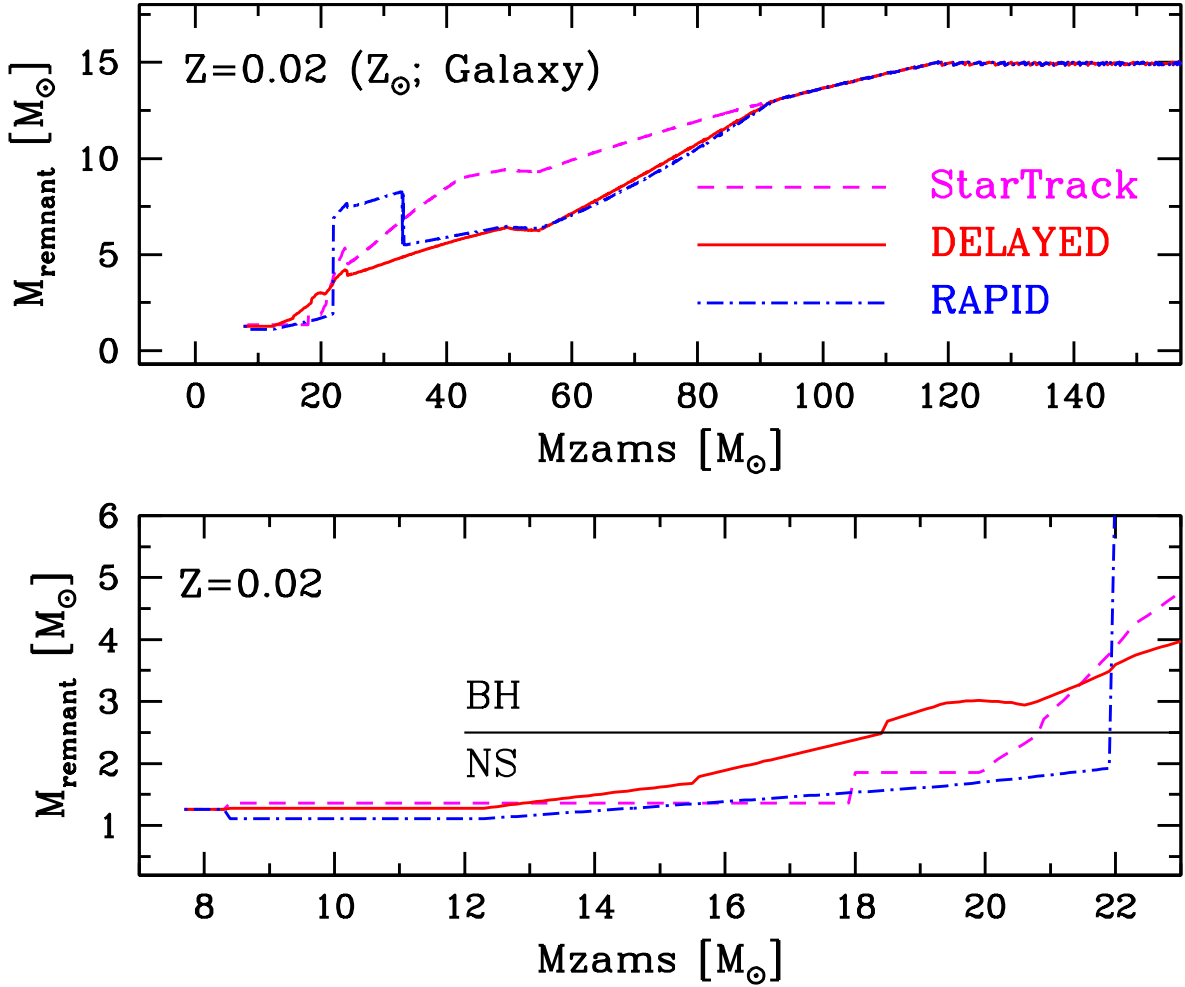}
\caption{Final compact object masses for the 3 presented calculation
  schemes for binary population synthesis in the function of initial
  mass for single star evolution (with solar metallicity and standard
  winds).  Top panel shows the full mass range, while bottom panel
  shows the mass range important for NS formation. The adopted
  transition mass from NS to BH is also indicated ($M_{\rm NS,max}=2.5
  \msun$).  }
\label{Mrem2}
\end{figure}

\begin{figure}
\plotone{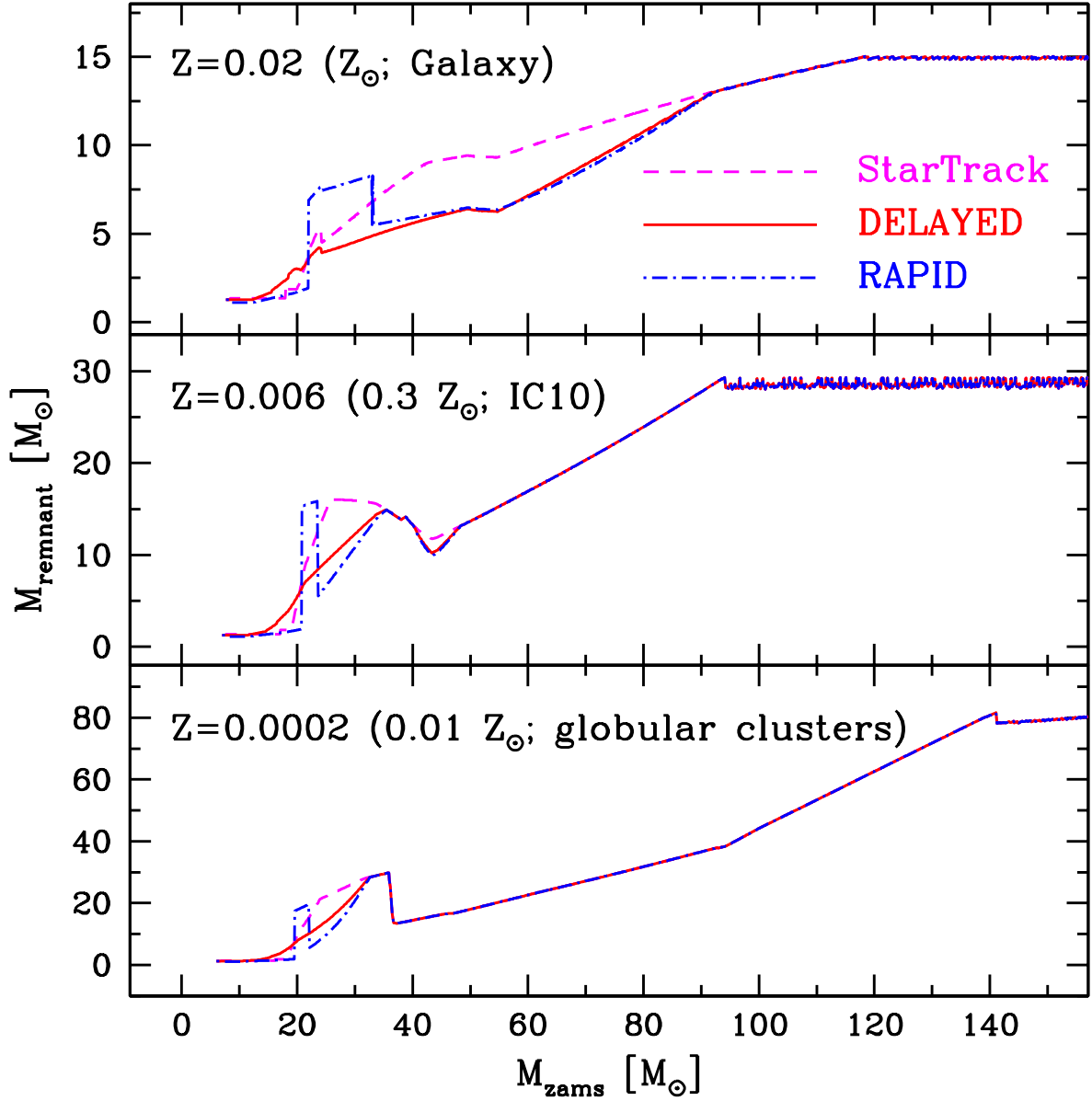}
\caption{Metallicity dependence of final compact object masses for
  single star evolution using our 3 calculation schemes for binary
  population synthesis.}
\label{Mrem1}
\end{figure}

\begin{figure}
\plotone{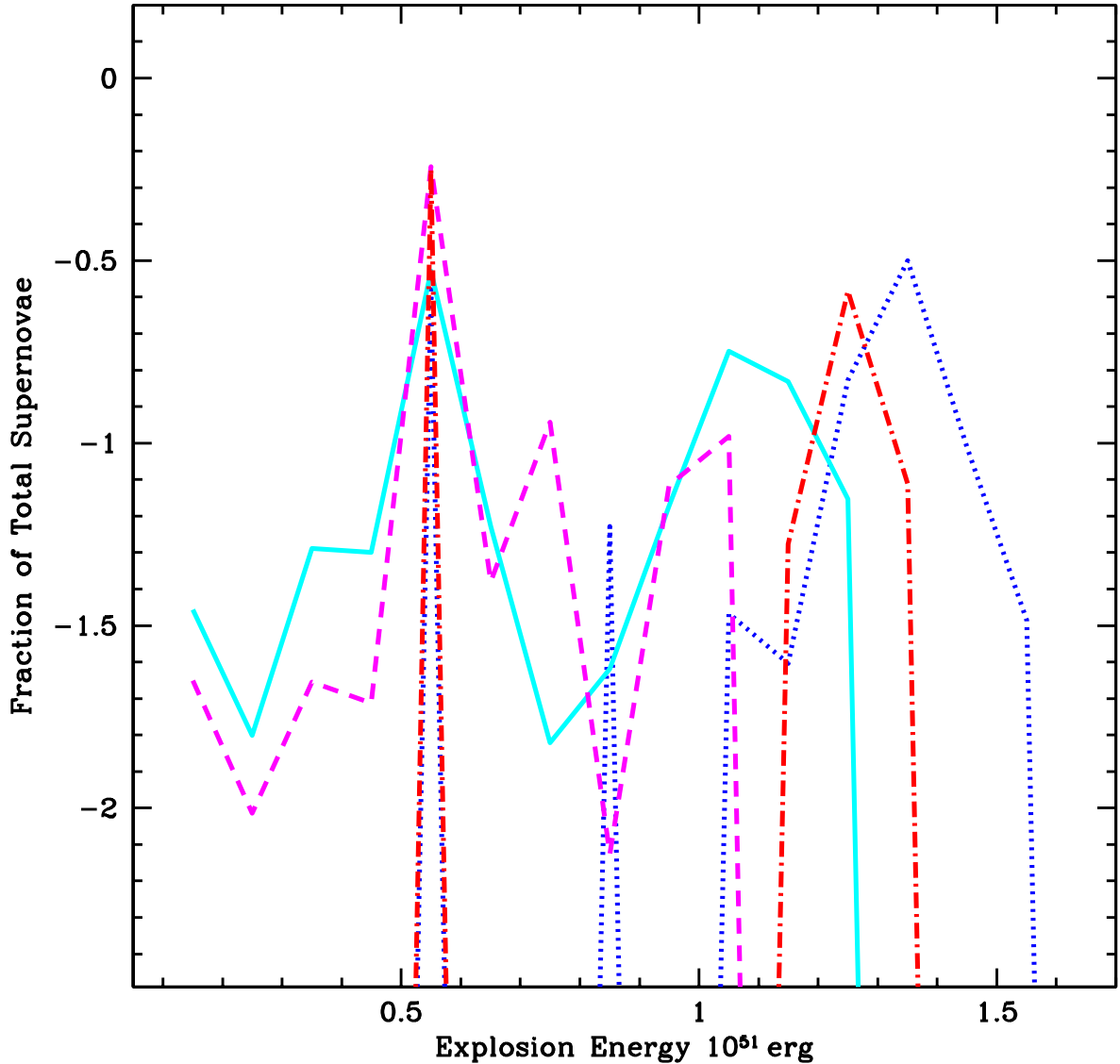}
\caption{Fraction of supernovae per 10$^{50}$\,erg energy bin.  We
  have assumed low mass supernovae all produce the same,
  $0.55\times10^{51}$\,erg, explosion, but it is likely that these
  stars produce a broader range of explosion energies.  Because the
  minimum mass for low-mass supernovae decreases at low
  metallicities, the fraction of supernovae produced by electron
  capture increases at high redshift.  The rapid explosion engine
  produces more energetic explosions.  Except for the low energy
  explosions from low-mass supernovae, this rapid explosion
  engine produces only explosions in excess of $10^{51}$\,erg.  The
  delayed explosion produces a broad range of explosion energies with 
  less than 50\% above $10^{51}$\,erg even at solar metallicity.}
\label{fig:distexpe}
\end{figure}

\begin{figure}
\includegraphics[scale=0.5,angle=-90]{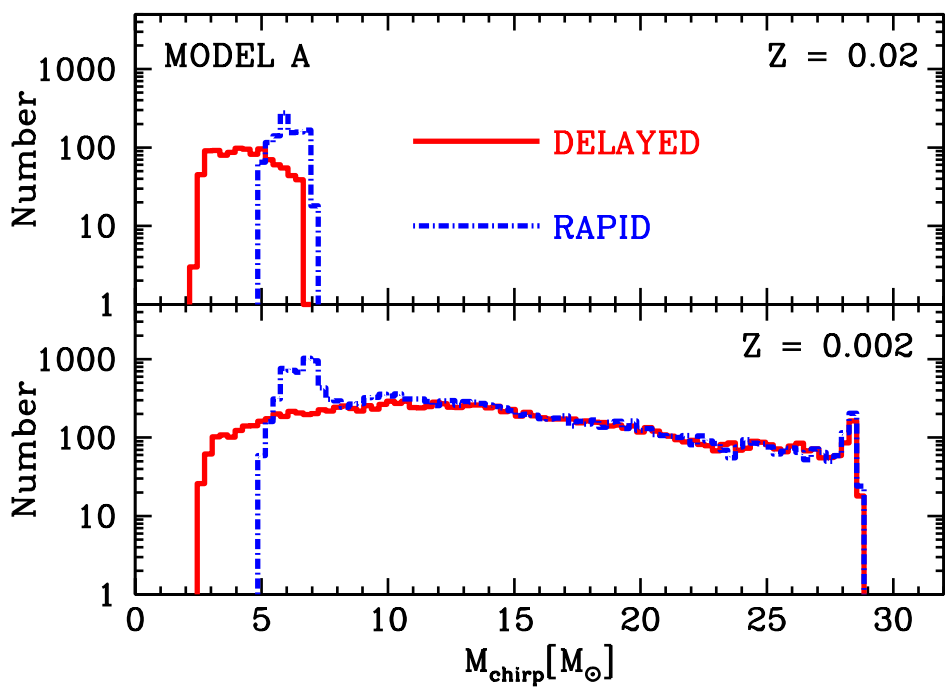}
\caption{Close BH-BH ($T_{delay}=t_{\rm evol} + t_{\rm merger}<10$ Gyr) chirp
mass distribution for the delayed (solid) and rapid (dot-dashed) supernova 
explosion models. The original {\tt StarTrack} model closely resembles the delayed model. Here 
we use model A for our common envelope evolution.}
\label{fig:chirpA}
\end{figure}

\begin{figure}
\includegraphics[scale=0.5,angle=-90]{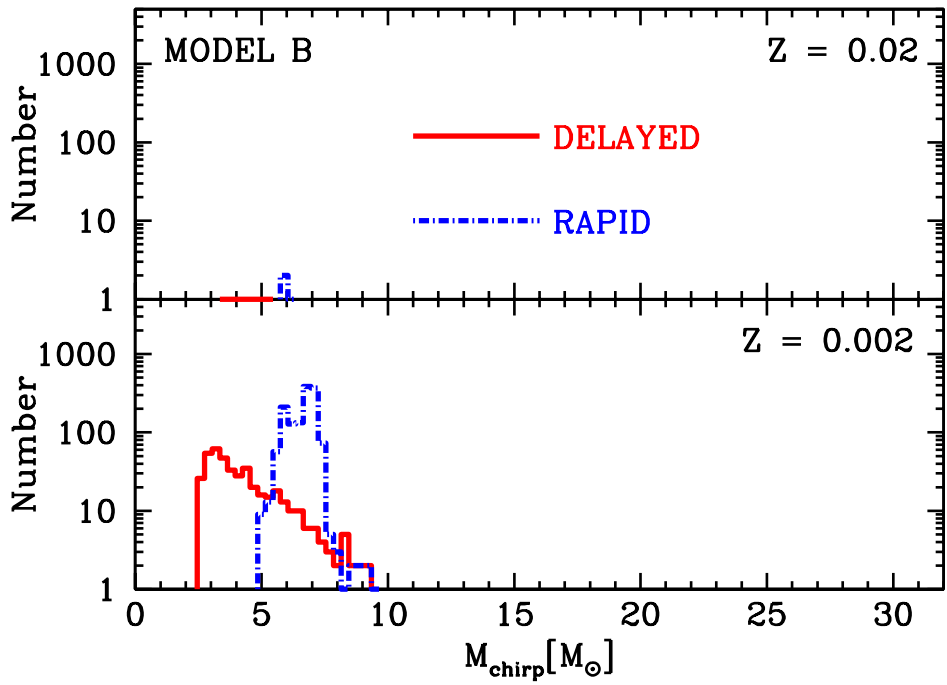}
\caption{The same as on Figure~\ref{fig:chirpA} but for model B of common envelope
evolution. The drastically reduced number of BH-BH binaries is due to very
frequent common envelope mergers in this model (e.g., Belczynski et al. 2007).}
\label{fig:chirpB}
\end{figure}

We predict the distribution of chirp masses for the two
supernova engine models (see Fig.\ref{fig:chirpA} and \ref{fig:chirpB}), and
find a striking difference. For the delayed model the chirp masses start from $M_{\rm chirp}
\approx 2.4(2.5)M_\odot$, whereas for the rapid model the lowest chirp
masses are $\approx 4.9(5.4)M_\odot$ for models A(B).
These very different chirp mass distributions are the result of the
differing initial-remnant mass relations presented in Figures~\ref{Mrem2}
and~\ref{Mrem1}. In the delayed model BHs start forming from $M_{\rm BH}=2.5 \msun$,
and their mass continuously increases as the initial mass of the progenitors increases.
The chirp mass distribution starts at a low value, $M_{\rm chirp}
\approx 2.4 M_\odot$ (the chirp mass of two $2.5 \msun$ black
holes is $2.2 \msun$). In contrast, in the rapid supernova engine model, there 
are no BHs with mass below $5 \msun$, and therefore the chirp mass
distribution starts at $M_{\rm chirp} \approx 5\msun$. The origin of this
remnant gap is described and referenced in Sec.~\ref{relations}.

In Table~\ref{numall} we present Galactic merger rates for our models
for BH-BH binaries.  
Rates are calculated for a Milky Way type galaxy (10 Gyr of continuous star
formation at a rate of $3.5 \mpy$), as described in Belczynski et al. (2007).
The change of rates with metallicity, and between our two models for common
envelope evolution, are fully
discussed in Belczynski et al. (2010b).
Here we note that both supernova models produce comparable numbers of
close BH-BH binaries, hence the similar merger rates in Table~\ref{numall}.
Therefore, observations of BH-BH binaries will illuminate the underlying
supernova engine only if the physical properties of 
the mergers can be extracted from the gravitational waves.  If the 
gap in the mass distribution from X-ray binaries is confirmed by gravitational 
wave observations of merging binaries, this gap will rule out delayed supernova 
mechanisms for most systems.

\acknowledgements
This project was funded in part under the auspices of the
U.S. Dept. of Energy, and supported by its contract W-7405-ENG-36 to
Los Alamos National Laboratory, and by a NASA grant SWIF03-0047.  K.B., 
M.D. and G.W. acknowledge support from MSHE grant N N203 404939.  V.K. 
acknowledges support from NSF grant AST-0908930 from this project.  
K.B., V.K., D.H. acknowledge partial support from the Aspen Center for 
Physics where part of this work was developed.

{}

\clearpage

\begin{deluxetable}{lccc}
\tablewidth{200pt}
\tablecaption{Galactic BH-BH Merger Rates [Myr$^{-1}$]\tablenotemark{a}}
\tablehead{& DELAYED & RAPID }

\startdata
$Z_\odot$               &  5.2 (0.02)  &  5.0 (0.01) \\
$0.1\ Z_\odot$          &  72.4  (4.0) &  87.4 (7.7) \\
\enddata
\label{numall}
\tablenotetext{a}{
Rates are given for model A(B).}
\end{deluxetable}

\clearpage

\clearpage

\clearpage

\clearpage

\clearpage

\clearpage

\clearpage

\clearpage

\clearpage

\clearpage

\clearpage

\clearpage

\clearpage

\clearpage

\end{document}